\documentclass[12pt]{article}
\usepackage{epsfig}
\newcommand{\epem}{\mathrm{e}^+\mathrm{e}^-} 
\newcommand{\WW}{\mathrm{W}^+\mathrm{W}^-} 
\newcommand{\qqbar}{\mathrm{q}\bar{\mathrm{q}}}
\newcommand{\nch}{\langle n_{\mathrm{ch}} \rangle}
\newcommand{\pto}{p_{\mathrm{t,out}}} 
\newcommand{\pti}{p_{\mathrm{t,in}}}
\newcommand{\Tm}{T_{\mathrm{minor}}} 
\newcommand{\Ecm}{E_{\mathrm{cm}}}
\newcommand{\Evis}{E_{\mathrm{vis}}}
\newcommand{\ksubT}{k_{\mathrm{T}}}
\newcommand{\ycut}{y_{\mathrm{cut}}}
\newcommand{\yupp}{y_{\mathrm{u}}}
\newcommand{\Pjet}{P_{\mathrm{jet}}} 
\newcommand{\Pg}{P_{\mathrm{g}}} 
\newcommand{\Ntj}{N_{3\mathrm{j}}}
\newcommand{\Preco}{P_{\mathrm{reco}}}
\newcommand{\kI}{k_{\mathrm{I}}}

\begin{document}
\setcounter{page}{0} 
\thispagestyle{empty}  
\begin{center}  {\large  EUROPEAN ORGANISATION FOR NUCLEAR RESEARCH (CERN)} \end{center} 
\vspace{5ex} 
\begin{flushright} CERN-PH-EP/2006-xxx  \\ 
                       21 April 2006  
\end{flushright} 
\vspace{10ex} 

\begin{center} {\LARGE \bf Test of Colour Reconnection Models \\  \vspace{2ex} 
                  using Three-Jet Events in Hadronic Z Decays} \\
\vspace{7ex} {\large  The ALEPH Collaboration}  \vspace{7ex} 
\end{center}  

\begin{abstract} 
Hadronic Z decays into three jets are used to test QCD models of colour reconnection (CR).
A sensitive quantity is the rate of gluon jets with a gap in the particle rapidity distribution  
and zero jet charge. Gluon jets are identified by either energy-ordering or by tagging two 
b-jets. The rates predicted by two string-based tunable CR models, one implemented in JETSET
(the GAL model), the other in ARIADNE, are too high and disfavoured by the data, whereas 
the rates from the corresponding non-CR standard versions of these generators are too low.
The data can be described by the GAL model assuming a small value for the $R_0$ parameter
in the range $0.01-0.02$. 
\end{abstract}  
\vspace{15ex}
\begin{center}  {\it Submitted to Eur. Phys. J. C}   \end{center}
        
\pagestyle{empty}
\newpage
\small
\newlength{\saveparskip}
\newlength{\savetextheight}
\newlength{\savetopmargin}
\newlength{\savetextwidth}
\newlength{\saveoddsidemargin}
\newlength{\savetopsep}
\setlength{\saveparskip}{\parskip}
\setlength{\savetextheight}{\textheight}
\setlength{\savetopmargin}{\topmargin}
\setlength{\savetextwidth}{\textwidth}
\setlength{\saveoddsidemargin}{\oddsidemargin}
\setlength{\savetopsep}{\topsep}
\setlength{\parskip}{0.0cm}
\setlength{\textheight}{25.0cm}
\setlength{\topmargin}{-1.5cm}
\setlength{\textwidth}{16 cm}
\setlength{\oddsidemargin}{-0.0cm}
\setlength{\topsep}{1mm}
\pretolerance=10000

\centerline{\large\bf The ALEPH Collaboration}
\footnotesize
\vspace{0.5cm}
{\raggedbottom
\begin{sloppypar}
\samepage\noindent
S.~Schael
\nopagebreak
\begin{center}
\parbox{15.5cm}{\sl\samepage
Physikalisches Institut das RWTH-Aachen, D-52056 Aachen, Germany}
\end{center}\end{sloppypar}
\vspace{2mm}
\begin{sloppypar}
\noindent
R.~Barate,
R.~Bruneli\`ere,
I.~De~Bonis,
D.~Decamp,
C.~Goy,
S.~J\'ez\'equel,
J.-P.~Lees,
F.~Martin,
E.~Merle,
\mbox{M.-N.~Minard},
B.~Pietrzyk,
B.~Trocm\'e
\nopagebreak
\begin{center}
\parbox{15.5cm}{\sl\samepage
Laboratoire de Physique des Particules (LAPP), IN$^{2}$P$^{3}$-CNRS,
F-74019 Annecy-le-Vieux Cedex, France}
\end{center}\end{sloppypar}
\vspace{2mm}
\begin{sloppypar}
\noindent
S.~Bravo,
M.P.~Casado,
M.~Chmeissani,
J.M.~Crespo,
E.~Fernandez,
M.~Fernandez-Bosman,
Ll.~Garrido,$^{15}$
M.~Martinez,
A.~Pacheco,
H.~Ruiz
\nopagebreak
\begin{center}
\parbox{15.5cm}{\sl\samepage
Institut de F\'{i}sica d'Altes Energies, Universitat Aut\`{o}noma
de Barcelona, E-08193 Bellaterra (Barcelona), Spain$^{7}$}
\end{center}\end{sloppypar}
\vspace{2mm}
\begin{sloppypar}
\noindent
A.~Colaleo,
D.~Creanza,
N.~De~Filippis,
M.~de~Palma,
G.~Iaselli,
G.~Maggi,
M.~Maggi,
S.~Nuzzo,
A.~Ranieri,
G.~Raso,$^{24}$
F.~Ruggieri,
G.~Selvaggi,
L.~Silvestris,
P.~Tempesta,
A.~Tricomi,$^{3}$
G.~Zito
\nopagebreak
\begin{center}
\parbox{15.5cm}{\sl\samepage
Dipartimento di Fisica, INFN Sezione di Bari, I-70126 Bari, Italy}
\end{center}\end{sloppypar}
\vspace{2mm}
\begin{sloppypar}
\noindent
X.~Huang,
J.~Lin,
Q. Ouyang,
T.~Wang,
Y.~Xie,
R.~Xu,
S.~Xue,
J.~Zhang,
L.~Zhang,
W.~Zhao
\nopagebreak
\begin{center}
\parbox{15.5cm}{\sl\samepage
Institute of High Energy Physics, Academia Sinica, Beijing, The People's
Republic of China$^{8}$}
\end{center}\end{sloppypar}
\vspace{2mm}
\begin{sloppypar}
\noindent
D.~Abbaneo,
T.~Barklow,$^{26}$
O.~Buchm\"uller,$^{26}$
M.~Cattaneo,
B.~Clerbaux,$^{23}$
H.~Drevermann,
R.W.~Forty,
M.~Frank,
F.~Gianotti,
J.B.~Hansen,
J.~Harvey,
D.E.~Hutchcroft,$^{30}$,
P.~Janot,
B.~Jost,
M.~Kado,$^{2}$
P.~Mato,
A.~Moutoussi,
F.~Ranjard,
L.~Rolandi,
D.~Schlatter,
F.~Teubert,
A.~Valassi,
I.~Videau
\nopagebreak
\begin{center}
\parbox{15.5cm}{\sl\samepage
European Laboratory for Particle Physics (CERN), CH-1211 Geneva 23,
Switzerland}
\end{center}\end{sloppypar}
\vspace{2mm}
\begin{sloppypar}
\noindent
F.~Badaud,
S.~Dessagne,
A.~Falvard,$^{20}$
D.~Fayolle,
P.~Gay,
J.~Jousset,
B.~Michel,
S.~Monteil,
D.~Pallin,
J.M.~Pascolo,
P.~Perret
\nopagebreak
\begin{center}
\parbox{15.5cm}{\sl\samepage
Laboratoire de Physique Corpusculaire, Universit\'e Blaise Pascal,
IN$^{2}$P$^{3}$-CNRS, Clermont-Ferrand, F-63177 Aubi\`{e}re, France}
\end{center}\end{sloppypar}
\vspace{2mm}
\begin{sloppypar}
\noindent
J.D.~Hansen,
J.R.~Hansen,
P.H.~Hansen,
A.C.~Kraan,
B.S.~Nilsson
\nopagebreak
\begin{center}
\parbox{15.5cm}{\sl\samepage
Niels Bohr Institute, 2100 Copenhagen, DK-Denmark$^{9}$}
\end{center}\end{sloppypar}
\vspace{2mm}
\begin{sloppypar}
\noindent
A.~Kyriakis,
C.~Markou,
E.~Simopoulou,
A.~Vayaki,
K.~Zachariadou
\nopagebreak
\begin{center}
\parbox{15.5cm}{\sl\samepage
Nuclear Research Center Demokritos (NRCD), GR-15310 Attiki, Greece}
\end{center}\end{sloppypar}
\vspace{2mm}
\begin{sloppypar}
\noindent
A.~Blondel,$^{12}$
\mbox{J.-C.~Brient},
F.~Machefert,
A.~Roug\'{e},
H.~Videau
\nopagebreak
\begin{center}
\parbox{15.5cm}{\sl\samepage
Laoratoire Leprince-Ringuet, Ecole
Polytechnique, IN$^{2}$P$^{3}$-CNRS, \mbox{F-91128} Palaiseau Cedex, France}
\end{center}\end{sloppypar}
\vspace{2mm}
\begin{sloppypar}
\noindent
V.~Ciulli,
E.~Focardi,
G.~Parrini
\nopagebreak
\begin{center}
\parbox{15.5cm}{\sl\samepage
Dipartimento di Fisica, Universit\`a di Firenze, INFN Sezione di Firenze,
I-50125 Firenze, Italy}
\end{center}\end{sloppypar}
\vspace{2mm}
\begin{sloppypar}
\noindent
A.~Antonelli,
M.~Antonelli,
G.~Bencivenni,
F.~Bossi,
G.~Capon,
F.~Cerutti,
V.~Chiarella,
P.~Laurelli,
G.~Mannocchi,$^{5}$
G.P.~Murtas,
L.~Passalacqua
\nopagebreak
\begin{center}
\parbox{15.5cm}{\sl\samepage
Laboratori Nazionali dell'INFN (LNF-INFN), I-00044 Frascati, Italy}
\end{center}\end{sloppypar}
\vspace{2mm}
\begin{sloppypar}
\noindent
J.~Kennedy,
J.G.~Lynch,
P.~Negus,
V.~O'Shea,
A.S.~Thompson
\nopagebreak
\begin{center}
\parbox{15.5cm}{\sl\samepage
Department of Physics and Astronomy, University of Glasgow, Glasgow G12
8QQ,United Kingdom$^{10}$}
\end{center}\end{sloppypar}
\vspace{2mm}
\begin{sloppypar}
\noindent
S.~Wasserbaech
\nopagebreak
\begin{center}
\parbox{15.5cm}{\sl\samepage
Utah Valley State College, Orem, UT 84058, U.S.A.}
\end{center}\end{sloppypar}
\vspace{2mm}
\begin{sloppypar}
\noindent
R.~Cavanaugh,$^{4}$
S.~Dhamotharan,$^{21}$
C.~Geweniger,
P.~Hanke,
V.~Hepp,
E.E.~Kluge,
A.~Putzer,
H.~Stenzel,
K.~Tittel,
M.~Wunsch$^{19}$
\nopagebreak
\begin{center}
\parbox{15.5cm}{\sl\samepage
Kirchhoff-Institut f\"ur Physik, Universit\"at Heidelberg, D-69120
Heidelberg, Germany$^{16}$}
\end{center}\end{sloppypar}
\vspace{2mm}
\begin{sloppypar}
\noindent
R.~Beuselinck,
W.~Cameron,
G.~Davies,
P.J.~Dornan,
M.~Girone,$^{1}$
N.~Marinelli,
J.~Nowell,
S.A.~Rutherford,
J.K.~Sedgbeer,
J.C.~Thompson,$^{14}$
R.~White
\nopagebreak
\begin{center}
\parbox{15.5cm}{\sl\samepage
Department of Physics, Imperial College, London SW7 2BZ,
United Kingdom$^{10}$}
\end{center}\end{sloppypar}
\vspace{2mm}
\begin{sloppypar}
\noindent
V.M.~Ghete,
P.~Girtler,
E.~Kneringer,
D.~Kuhn,
G.~Rudolph
\nopagebreak
\begin{center}
\parbox{15.5cm}{\sl\samepage
Institut f\"ur Experimentalphysik, Universit\"at Innsbruck, A-6020
Innsbruck, Austria$^{18}$}
\end{center}\end{sloppypar}
\vspace{2mm}
\begin{sloppypar}
\noindent
E.~Bouhova-Thacker,
C.K.~Bowdery,
D.P.~Clarke,
G.~Ellis,
A.J.~Finch,
F.~Foster,
G.~Hughes,
R.W.L.~Jones,
M.R.~Pearson,
N.A.~Robertson,
M.~Smizanska
\nopagebreak
\begin{center}
\parbox{15.5cm}{\sl\samepage
Department of Physics, University of Lancaster, Lancaster LA1 4YB,
United Kingdom$^{10}$}
\end{center}\end{sloppypar}
\vspace{2mm}
\begin{sloppypar}
\noindent
O.~van~der~Aa,
C.~Delaere,$^{28}$
G.Leibenguth,$^{31}$
V.~Lemaitre$^{29}$
\nopagebreak
\begin{center}
\parbox{15.5cm}{\sl\samepage
Institut de Physique Nucl\'eaire, D\'epartement de Physique, Universit\'e Catholique de Louvain, 1348 Louvain-la-Neuve, Belgium}
\end{center}\end{sloppypar}
\vspace{2mm}
\begin{sloppypar}
\noindent
U.~Blumenschein,
F.~H\"olldorfer,
K.~Jakobs,
F.~Kayser,
A.-S.~M\"uller,
B.~Renk,
H.-G.~Sander,
S.~Schmeling,
H.~Wachsmuth,
C.~Zeitnitz,
T.~Ziegler
\nopagebreak
\begin{center}
\parbox{15.5cm}{\sl\samepage
Institut f\"ur Physik, Universit\"at Mainz, D-55099 Mainz, Germany$^{16}$}
\end{center}\end{sloppypar}
\vspace{2mm}
\begin{sloppypar}
\noindent
A.~Bonissent,
P.~Coyle,
C.~Curtil,
A.~Ealet,
D.~Fouchez,
P.~Payre,
A.~Tilquin
\nopagebreak
\begin{center}
\parbox{15.5cm}{\sl\samepage
Centre de Physique des Particules de Marseille, Univ M\'editerran\'ee,
IN$^{2}$P$^{3}$-CNRS, F-13288 Marseille, France}
\end{center}\end{sloppypar}
\vspace{2mm}
\begin{sloppypar}
\noindent
F.~Ragusa
\nopagebreak
\begin{center}
\parbox{15.5cm}{\sl\samepage
Dipartimento di Fisica, Universit\`a di Milano e INFN Sezione di
Milano, I-20133 Milano, Italy.}
\end{center}\end{sloppypar}
\vspace{2mm}
\begin{sloppypar}
\noindent
A.~David,
H.~Dietl,$^{32}$
G.~Ganis,$^{27}$
K.~H\"uttmann,
G.~L\"utjens,
W.~M\"anner$^{32}$,
\mbox{H.-G.~Moser},
R.~Settles,
M.~Villegas,
G.~Wolf
\nopagebreak
\begin{center}
\parbox{15.5cm}{\sl\samepage
Max-Planck-Institut f\"ur Physik, Werner-Heisenberg-Institut,
D-80805 M\"unchen, Germany\footnotemark[16]}
\end{center}\end{sloppypar}
\vspace{2mm}
\begin{sloppypar}
\noindent
J.~Boucrot,
O.~Callot,
M.~Davier,
L.~Duflot,
\mbox{J.-F.~Grivaz},
Ph.~Heusse,
A.~Jacholkowska,$^{6}$
L.~Serin,
\mbox{J.-J.~Veillet}
\nopagebreak
\begin{center}
\parbox{15.5cm}{\sl\samepage
Laboratoire de l'Acc\'el\'erateur Lin\'eaire, Universit\'e de Paris-Sud,
IN$^{2}$P$^{3}$-CNRS, F-91898 Orsay Cedex, France}
\end{center}\end{sloppypar}
\vspace{2mm}
\begin{sloppypar}
\noindent
P.~Azzurri, 
G.~Bagliesi,
T.~Boccali,
L.~Fo\`a,
A.~Giammanco,
A.~Giassi,
F.~Ligabue,
A.~Messineo,
F.~Palla,
G.~Sanguinetti,
A.~Sciab\`a,
G.~Sguazzoni,
P.~Spagnolo,
R.~Tenchini,
A.~Venturi,
P.G.~Verdini
\samepage
\begin{center}
\parbox{15.5cm}{\sl\samepage
Dipartimento di Fisica dell'Universit\`a, INFN Sezione di Pisa,
e Scuola Normale Superiore, I-56010 Pisa, Italy}
\end{center}\end{sloppypar}
\vspace{2mm}
\begin{sloppypar}
\noindent
O.~Awunor,
G.A.~Blair,
G.~Cowan,
A.~Garcia-Bellido,
M.G.~Green,
T.~Medcalf,$^{25}$
A.~Misiejuk,
J.A.~Strong,
P.~Teixeira-Dias
\nopagebreak
\begin{center}
\parbox{15.5cm}{\sl\samepage
Department of Physics, Royal Holloway \& Bedford New College,
University of London, Egham, Surrey TW20 OEX, United Kingdom$^{10}$}
\end{center}\end{sloppypar}
\vspace{2mm}
\begin{sloppypar}
\noindent
R.W.~Clifft,
T.R.~Edgecock,
P.R.~Norton,
I.R.~Tomalin,
J.J.~Ward
\nopagebreak
\begin{center}
\parbox{15.5cm}{\sl\samepage
Particle Physics Dept., Rutherford Appleton Laboratory,
Chilton, Didcot, Oxon OX11 OQX, United Kingdom$^{10}$}
\end{center}\end{sloppypar}
\vspace{2mm}
\begin{sloppypar}
\noindent
\mbox{B.~Bloch-Devaux},
D.~Boumediene,
P.~Colas,
B.~Fabbro,
E.~Lan\c{c}on,
\mbox{M.-C.~Lemaire},
E.~Locci,
P.~Perez,
J.~Rander,
B.~Tuchming,
B.~Vallage
\nopagebreak
\begin{center}
\parbox{15.5cm}{\sl\samepage
CEA, DAPNIA/Service de Physique des Particules,
CE-Saclay, F-91191 Gif-sur-Yvette Cedex, France$^{17}$}
\end{center}\end{sloppypar}
\vspace{2mm}
\begin{sloppypar}
\noindent
A.M.~Litke,
G.~Taylor
\nopagebreak
\begin{center}
\parbox{15.5cm}{\sl\samepage
Institute for Particle Physics, University of California at
Santa Cruz, Santa Cruz, CA 95064, USA$^{22}$}
\end{center}\end{sloppypar}
\vspace{2mm}
\begin{sloppypar}
\noindent
C.N.~Booth,
S.~Cartwright,
F.~Combley,$^{25}$
P.N.~Hodgson,
M.~Lehto,
L.F.~Thompson
\nopagebreak
\begin{center}
\parbox{15.5cm}{\sl\samepage
Department of Physics, University of Sheffield, Sheffield S3 7RH,
United Kingdom$^{10}$}
\end{center}\end{sloppypar}
\vspace{2mm}
\begin{sloppypar}
\noindent
A.~B\"ohrer,
S.~Brandt,
C.~Grupen,
J.~Hess,
A.~Ngac,
G.~Prange
\nopagebreak
\begin{center}
\parbox{15.5cm}{\sl\samepage
Fachbereich Physik, Universit\"at Siegen, D-57068 Siegen, Germany$^{16}$}
\end{center}\end{sloppypar}
\vspace{2mm}
\begin{sloppypar}
\noindent
C.~Borean,
G.~Giannini
\nopagebreak
\begin{center}
\parbox{15.5cm}{\sl\samepage
Dipartimento di Fisica, Universit\`a di Trieste e INFN Sezione di Trieste,
I-34127 Trieste, Italy}
\end{center}\end{sloppypar}
\vspace{2mm}
\begin{sloppypar}
\noindent
H.~He,
J.~Putz,
J.~Rothberg
\nopagebreak
\begin{center}
\parbox{15.5cm}{\sl\samepage
Experimental Elementary Particle Physics, University of Washington, Seattle,
WA 98195 U.S.A.}
\end{center}\end{sloppypar}
\vspace{2mm}
\begin{sloppypar}
\noindent
S.R.~Armstrong,
K.~Berkelman,
K.~Cranmer,
D.P.S.~Ferguson,
Y.~Gao,$^{13}$
S.~Gonz\'{a}lez,
O.J.~Hayes,
H.~Hu,
S.~Jin,
J.~Kile,
P.A.~McNamara III,
J.~Nielsen,
Y.B.~Pan,
\mbox{J.H.~von~Wimmersperg-Toeller}, 
W.~Wiedenmann,
J.~Wu,
Sau~Lan~Wu,
X.~Wu,
G.~Zobernig
\nopagebreak
\begin{center}
\parbox{15.5cm}{\sl\samepage
Department of Physics, University of Wisconsin, Madison, WI 53706,
USA$^{11}$}
\end{center}\end{sloppypar}
\vspace{2mm}
\begin{sloppypar}
\noindent
G.~Dissertori
\nopagebreak
\begin{center}
\parbox{15.5cm}{\sl\samepage
Institute for Particle Physics, ETH H\"onggerberg, 8093 Z\"urich,
Switzerland.}
\end{center}\end{sloppypar}
}
\footnotetext[1]{Also at CERN, 1211 Geneva 23, Switzerland.}
\footnotetext[2]{Now at Fermilab, PO Box 500, MS 352, Batavia, IL 60510, USA}
\footnotetext[3]{Also at Dipartimento di Fisica di Catania and INFN Sezione di
 Catania, 95129 Catania, Italy.}
\footnotetext[4]{Now at University of Florida, Department of Physics, Gainesville, Florida 32611-8440, USA}
\footnotetext[5]{Also IFSI sezione di Torino, INAF, Italy.}
\footnotetext[6]{Also at Groupe d'Astroparticules de Montpellier, Universit\'{e} de Montpellier II, 34095, Montpellier, France.}
\footnotetext[7]{Supported by CICYT, Spain.}
\footnotetext[8]{Supported by the National Science Foundation of China.}
\footnotetext[9]{Supported by the Danish Natural Science Research Council.}
\footnotetext[10]{Supported by the UK Particle Physics and Astronomy Research
Council.}
\footnotetext[11]{Supported by the US Department of Energy, grant
DE-FG0295-ER40896.}
\footnotetext[12]{Now at Departement de Physique Corpusculaire, Universit\'e de
Gen\`eve, 1211 Gen\`eve 4, Switzerland.}
\footnotetext[13]{Also at Department of Physics, Tsinghua University, Beijing, The People's Republic of China.}
\footnotetext[14]{Supported by the Leverhulme Trust.}
\footnotetext[15]{Permanent address: Universitat de Barcelona, 08208 Barcelona,
Spain.}
\footnotetext[16]{Supported by Bundesministerium f\"ur Bildung
und Forschung, Germany.}
\footnotetext[17]{Supported by the Direction des Sciences de la
Mati\`ere, C.E.A.}
\footnotetext[18]{Supported by the Austrian Ministry for Science and Transport.}
\footnotetext[19]{Now at SAP AG, 69185 Walldorf, Germany}
\footnotetext[20]{Now at Groupe d' Astroparticules de Montpellier, Universit\'e de Montpellier II, 34095 Montpellier, France.}
\footnotetext[21]{Now at BNP Paribas, 60325 Frankfurt am Mainz, Germany}
\footnotetext[22]{Supported by the US Department of Energy,
grant DE-FG03-92ER40689.}
\footnotetext[23]{Now at Institut Inter-universitaire des hautes Energies (IIHE), CP 230, Universit\'{e} Libre de Bruxelles, 1050 Bruxelles, Belgique}
\footnotetext[24]{Now at Dipartimento di Fisica e Tecnologie Relative, Universit\`a di Palermo, Palermo, Italy.}
\footnotetext[25]{Deceased.}
\footnotetext[26]{Now at SLAC, Stanford, CA 94309, U.S.A}
\footnotetext[27]{Now at CERN, 1211 Geneva 23, Switzerland}
\footnotetext[28]{Research Fellow of the Belgium FNRS}
\footnotetext[29]{Research Associate of the Belgium FNRS} 
\footnotetext[30]{Now at Liverpool University, Liverpool L69 7ZE, United Kingdom} 
\footnotetext[31]{Supported by the Federal Office for Scientific, Technical and Cultural Affairs through
the Interuniversity Attraction Pole P5/27} 
\footnotetext[32]{Now at Henryk Niewodnicznski Institute of Nuclear Physics, Polish Academy of Sciences, Cracow, Poland}   
\setlength{\parskip}{\saveparskip}
\setlength{\textheight}{\savetextheight}
\setlength{\topmargin}{\savetopmargin}
\setlength{\textwidth}{\savetextwidth}
\setlength{\oddsidemargin}{\saveoddsidemargin}
\setlength{\topsep}{\savetopsep}
\normalsize
\newpage
\pagestyle{plain}
\setcounter{page}{1}

\section{Introduction} 
A description of the hadronisation of a multiparton system requires specification of the colour
connections among the partons. These can be modified by higher-order interference or 
non-perturbative effects in QCD, a phenomenon commonly called Colour Reconnection (CR) \cite{CR1}.
These effects can only be studied within the framework of specific models.     
The present interest in CR arises from the precise measurement of the W boson mass in 
$\WW \rightarrow \qqbar \qqbar$ events in $\epem$ collisions at LEP2 energies.   
In this fully hadronic final state possible CR effects among the decay products  
of the two W bosons contribute the largest systematic uncertainty to $m_{\mathrm{W}}$ \cite{LEP-MW}.
Direct information on CR in this channel is obtained from inter-jet particle flow studies 
\cite{pf_L3,pf_OPAL} and/or from the W mass itself by comparing different jet algorithms   
\cite{mw_OPAL,mw_ALEPH}. Due to limited statistics only extreme models can be excluded.  

CR effects might also appear within a colour singlet system like the one produced in the reaction  
$\epem \rightarrow \qqbar$(+gluons) on the Z resonance at LEP1 where high statistics data are     
available. To search for CR effects one has to look for hadronic variables which are sensitive to  
the colour flow in an event. Following a proposal in Ref.\ \cite{CR2}, the OPAL collaboration has
shown that gluon jets with a rapidity gap and zero jet charge, identified in events of the type 
Z $\rightarrow 3$ jets, provide a sensitive means to search for CR effects, and concluded
that two string-based CR models are disfavoured by their data \cite{ygap_OPAL,ydis_OPAL}. 
The L3 collaboration, employing angular gaps in the inter-jet regions of symmetric three-jet
events, arrived at a similar conclusion \cite{ygap_L3}.
By comparing quark with gluon jets, the DELPHI collaboration found that also the standard 
colour string model (without CR) cannot adequately describe the fraction of neutral
gluon jets with a rapidity gap \cite{ql_DELPHI}. They suggest a contribution either from a colour
octet neutralisation mechanism \cite{glueball} or, alternatively, from colour reconnection.    

In the present paper the variables proposed in \cite{ygap_OPAL} are used both to test CR models 
and to investigate the discrepancy reported in \cite{ql_DELPHI}. The data were collected by the 
ALEPH detector at LEP1. The data are compared to QCD Monte Carlo model calculations with and
without implementation of CR and with parameters tuned to global properties of 
hadronic Z decays. Gluon jets from three-jet events are identified by either energy-ordering
or by anti-b tagging. The rate of neutral jets is studied as a function of the rapidity gap
size. Quark jets from the same events are used for purposes of comparison. 
The influence of Bose-Einstein correlations is also investigated.

\section{QCD models for colour reconnection} 
Some basic properties of the non-perturbative CR models to be discussed in this paper are  
listed in Table \ref{crmod}. 
\begin{table}   \begin{center}   
\caption{\label{crmod} \small Properties of colour reconnection models.}  \vspace{1ex}                      
\begin{tabular}{|ll|ll|c|}   \hline
model   &  criterion of    &  free     &       &  effect in  \\
        & reconnection     & parameter & value & Z$\rightarrow \qqbar$ \\  \hline
\hline  
SKI     & space-time overlap        & $\kI$    & 0.65  & not          \\
        & of flux tubes             &          &       & implemented  \\  \hline  
SKII    & crossing of               &   -      &  -    &   No         \\
        & vortex lines              &          &       &              \\  \hline 
ARIADNE & reduce total              & $\Preco$ & 1/9   &   Yes        \\  
AR1     & string length $\lambda$   &          &       &              \\  \hline 
GAL     & reduce total              & $R_0$    &  0.1  &   Yes        \\
(Rathsman)& string area $A$         &          &       &              \\  \hline  
HERWIG  & reduce cluster size       & $\Preco$ & 1/9   &  Yes         \\
        & in space-time             &          &       &              \\  \hline  	
\end{tabular}  
\end{center}  \end{table}  
The first detailed theoretical study of CR was carried out by Sj\"ostrand and Khoze (SK) \cite{SK}    
in the context of a possible cross-talk among the hadronically decaying W bosons in W pair 
production and its effect on the W mass measurement. The perturbative QCD 
contribution from one-gluon exchange was found to be negligible, but there could be a sizeable  
non-perturbative contribution. Using the leading-log approximation (LLA), 
each $\qqbar$ system is evolved into a parton shower which, in the 
large-$N_{\mathrm{C}}$ limit  (where $N_{\mathrm{C}}$ is the number of colours), 
determines a sequence of colour-connected partons (quarks and gluons) 
which is used to draw the colour string.  
The authors developed a non-perturbative reconnection model, implemented \cite{PY61} in the PYTHIA 
generator version 6, based on the space-time overlap of the colour strings. 
The model variants SKI and SKII may be considered as two extreme descriptions of the colour topology.
In SKI, strings are assumed to be extended flux tubes.
The probability to reconnect two strings is related to the overlap integral $I$ and is given by  
                         \[  \Preco = 1-\exp(-\kI \cdot I)  \]
where $\kI$ is a free parameter. 
In the SKII model the information is contained in thin vortex lines. 
Reconnection is assumed to take place with unit probability if they cross each other.  
The fact that the fraction of reconnected events is predicted in model SKII can be used to fix the 
parameter $\kI$ such that this fraction is the same in the two models ($\kI=0.65$). In the SKI
model, CR effects would in principle also be possible within colour singlet (CS) systems like W  
or Z decays, but they have not been implemented and are thus not testable. It is remarkable that 
the SKII model predicts that such reconnections should not occur at all within singlet systems 
since the partons emerge from a single vertex.    

The ARIADNE generator, version 4,  which is based on the colour dipole model, provides options for  
colour reconnection \cite{AR-CR}. Considering all pairs of non-adjacent dipoles, reconnections  
take place with probability $1/N_{\mathrm{C}}^2$ if the total string length $\lambda$ decreases. 
The $\lambda$ measure is defined from the four-momenta of $n$ colour-ordered partons: 

    \[  \lambda=\sum_{i=1}^{n-1} \log((p_i + p_{i+1})^2/m_0^2)   \]             with $m_0=1$ GeV.  
In the program, $N_{\mathrm{C}}^2$ may be considered a free parameter, normally set to 9.   
The option AR1 used in this paper enables reconnections only within colour singlet systems, 
whereas AR2 and AR3 are foreseen for reconnections also between CS systems like W pairs.  

The model due to Rathsman \cite{GAL} is based on the so-called generalized area law (GAL). 
The total area of a string is the sum of the areas of its pieces: 

    \[  A=\sum_{i=1}^{n-1} ((p_i + p_{i+1})^2 - (m_i + m_{i+1})^2)    \] 
At the end of the parton shower, pairs of string pieces are allowed to reconnect  
with probability
\begin{equation}  
  \Preco=R_0 (1-\exp(-b \Delta A)), \:\: \mbox{where} \:\: \Delta A = \max(0, A-A_{\mathrm{reco}}).  
\end{equation} 
$A_{\mathrm{reco}}$ is the area after a colour rearrangement and the positive parameter $b$ 
is one of the two parameters of the Lund symmetric fragmentation function. The phenomenological   
parameter $R_0$ should be of order $1/N_{\mathrm{C}}^2$. Its value has been determined
by the author to be 0.10 from a comparison of the model to HERA data on the diffractive  
structure function of the proton. 
The Rathsman program is interfaced with PYTHIA(JETSET) version 5.7 \cite{PY57}.   

The HERWIG generator, version 6.1 \cite{HW-CR} or higher, offers a quite different concept for
colour reconnection based on the space-time structure of an event at the partonic level. The 
relevant quantity is the cluster size defined as the Lorentz invariant space-time distance between
the calculated production points of the quark and the antiquark forming a cluster. 
A reconnection among pairs of non-adjacent clusters is performed with probability $\Preco=1/9$ 
if the sum of the squared cluster sizes is lowered.  
On average, colour reconnection in this model leads to higher cluster masses.

\section{QCD model tuning} 
Multi-hadronic final states are described by QCD Monte Carlo generators  
which contain a number of free parameters. These have to be determined from fits to data. 
This has been done extensively using event-shape, charged particle momentum and identified hadron 
momentum distributions measured by ALEPH in inclusive Z decays in order to check the overall   
description of the data and to obtain optimal values for the free parameters \cite{QCDmega}. 
This section describes the tuning of the QCD generators when including colour reconnection.
One of the effects of CR is that the average particle multiplicity changes slightly. 
For example, the average charged particle multiplicity, $\nch$, changes by --2\%, --1\% and +1\% 
for the GAL, AR1 and HW-CR models, respectively, as compared to the non-reconnected  
versions without changing any fragmentation parameters.  
Therefore the most important fragmentation parameters have been re-tuned for the colour   
reconnected versions of JETSET(GAL), ARIADNE and HERWIG. 
The tuning of HERWIG is described in more detail in \cite{GR_Mor00}.  
The best fit values of the parameters  are given in  
Tables \ref{JStune}--\ref{HWtune} and the $\chi^2$ values are summarized in Table \ref{gfchi2}.  
The results from the standard, non-reconnected versions are also included in the tables. For JETSET  
and ARIADNE, the string fragmentation parameters controlling the spin, flavour and type of the         
produced hadrons and the heavy quark fragmentation parameters $\epsilon_c, \: \epsilon_b$ are taken 
from Table 8 of  \cite{QCDmega}. The tuning of JETSET including a simulation of Bose-Einstein
correlations (model BE$_{32}$) is described in \cite{BEWW_ALEPH}.        
Also given in Tables \ref{JStune}--\ref{HWtune} are the predicted charged particle multiplicities in
hadronic events to be compared to the measured value $20.91\pm0.22$ \cite{QCDmega}.   

The multidimensional fitting method is described in  \cite{QCDmega}.  
The set of distributions used in the fits is listed in Table \ref{gfchi2}.
The experimental distributions of these variables, corrected for detector and ISR effects, 
are those presented in \cite{QCDmega} based on data collected in 1992.   
They are combined with data from the Z peak running in 1993 
in order to increase the statistics to about 1 million hadronic events.
The systematic errors are those evaluated for the 1992 sample. Since the total errors are dominated 
by systematics and since the systematic uncertainties are only rough estimates,  
the $\chi^2$ values given in Table \ref{gfchi2} should only be considered as a relative measure of 
the fit quality.   
In performing the fits of JETSET(+GAL) and ARIADNE, certain regions are excluded where these models 
clearly deviate from the data (the tails of the out-of-plane quantities 
$A \ge 0.06, \:\: \Tm \ge 0.20, \:\: \pto \ge 0.7$ GeV 
and the very low $x_p \le 0.014$).
For the HERWIG fits the momentum spectra of non-strange and strange mesons as 
well as baryons are included in the set of distributions.   
     
\begin{table}  \begin{center}  
\caption{\label{JStune} \small Tuned JETSET 7.4 parameters without and with inclusion of CR. 
Azimuthal isotropy in the parton shower is assumed (MSTJ(46)=0), since this option gives a better 
fit to the data. 
Any parameter given without error is fixed during the fit procedure.
The fraction of reconnected events is given in the last line.}    \vspace{1ex}
\begin{tabular}{|ll|l|l|l|}   \hline
  parameter      & MC name  & JETSET          &  +GAL           &  +GAL         \\  \hline \hline
 $R_0$           & PARP(188)&  -              & $0.039\pm0.011$ &   0.10           \\ \hline
 $\Lambda$ (GeV) & PARJ(81) & $0.312\pm0.004$ & $0.307\pm0.006$ & $0.306\pm0.006$  \\
 $Q_0$ (GeV)     & PARJ(82) & $1.50\pm0.07$   & $1.57\pm0.08$   & $1.79\pm0.04$    \\
 $\sigma$ (GeV)  & PARJ(21) & $0.365\pm0.009$ & $0.364\pm0.009$ & $0.362\pm0.009$  \\
 $a$             & PARJ(41) & 0.40            & 0.40            & 0.40             \\
 $b$ (GeV)$^{-2}$& PARJ(42) & $0.900\pm0.018$ & $0.815\pm0.026$ & $0.724\pm0.014$  \\ \hline
 $\nch$          &          & 20.64           &  20.65          &  20.67           \\ 
 f(reco)         &          & -               &   0.10          &   0.18           \\ \hline 
\end{tabular}      

\caption{\label{ARtune} \small Tuned ARIADNE 4.08 parameters without and with inclusion of CR.   
Any parameter given without error is fixed during the fit procedure.
The fraction of reconnected events is given in the last line.}    \vspace{1ex} 
\begin{tabular}{|ll|l|l|}   \hline
 parameter         & MC name     &  AR0            &  AR1              \\  \hline  \hline
 CR option         & MSTA(35)    &  0              & 1                 \\ 
 $N_C^2$           & PARA(26)    &  -              & 9                 \\  \hline
 $\Lambda$ (GeV)   & PARA(1)     & $0.229\pm0.003$ & $0.230\pm0.003$   \\
 $p_{\mathrm{T,min}}$ (GeV) 
                   & PARA(3)     & $0.79\pm0.05$   & $0.79\pm0.02$     \\
 $\sigma$ (GeV)    & PARJ(21)    & $0.358\pm0.009$ & $0.353\pm0.009$   \\
 $a$               & PARJ(41)    & 0.40            & 0.40              \\
 $b$ (GeV)$^{-2}$  & PARJ(42)    & $0.825\pm0.024$ & $0.758\pm0.015$   \\  \hline
 $\nch$            &             & 20.58           & 20.61             \\  
f(reco)            &             & -               & 0.15              \\  \hline 
\end{tabular}   
  
\caption{\label{HWtune}  \small 
Tuned HERWIG 6.1 parameters without and with inclusion of CR. Two cluster model 
parameters, CLSMR and PSPLT, have been made flavour-dependent in order to better describe the 
measured B-meson fragmentation function. The D-wave meson multiplets are switched off. 
Any parameter given without error is fixed during the fit procedure.
The fraction of reconnected events is given in the last line.}    \vspace{1ex} 
\begin{tabular}{|ll|l|l|}   \hline
 parameter              & MC name     &   HW0           &  HW-CR          \\ \hline \hline
 $\Preco$               & PRECO       &  0              &  1/9            \\ 
 min. virtuality (GeV$^2)$ & VMIN2    &  -              &  0.1            \\  \hline    
 $\Lambda$ (GeV)        & QCDLAM      & $0.190\pm0.005$ & $0.187\pm0.005$ \\
 gluon mass (GeV)       & RMASS(13)   & $0.77\pm0.01$   & $0.79\pm0.01$   \\
 max. cluster mass (GeV)& CLMAX       & $3.39\pm0.08$   & $3.40\pm0.08$   \\  
 angular smearing, dusc & CLSMR(1)    & $0.59\pm0.03$   & $0.66\pm0.04$   \\
 angular smearing, b    & CLSMR(2)    & 0               & 0               \\
 power in cluster       &             &                 &                 \\  
 splitting, dusc        & PSPLT(1)    & $0.945\pm0.018$ & $0.886\pm0.017$ \\
 power in cluster       &             &                 &                 \\ 
 splitting, b           & PSPLT(2)    & 0.33            & 0.32            \\
 decuplet baryon weight & DECWT       & $0.71\pm0.06$   & $0.70\pm0.06$   \\ \hline
 $\nch$                 &             & 20.96           & 20.98           \\  
 f(reco)                &             & -               & 0.08            \\ \hline 
\end{tabular} 
\end{center} \end{table}                 

\begin{table}  \begin{center}  
\caption{\label{gfchi2} \small $\chi^2$ values for the event-shape and charged particle momentum 
 distributions used in the global fits. The total $\chi^2$ values for the fits using restricted 
 regions are given in the last line.}    \vspace{1ex}
\begin{tabular}{|lr|rrr|rr|rr|}   \hline
                & model  &JETSET &  GAL  &  GAL  &   AR0 &   AR1 &   HW0 & HW-CR  \\
              & CR par.  &  0    &  0.04 &  0.10 &   0   &  1/9  &    0  &   1/9  \\
distribution    & bins   &       &       &       &       &       &       &        \\   \hline
sphericity, $S$ &    23  &    5  &    7  &   17  &   27  &   24  &  107  &   127  \\ 
aplanarity, $A$ &    16  &   91  &  117  &  132  &   44  &   60  &   76  &    93  \\
thrust, $T$     &    21  &   51  &   21  &    8  &   15  &   12  &  378  &   411  \\ 
minor, $\Tm$    &    18  &   76  &   99  &  130  &   64  &   81  &  162  &   200  \\ 
Feynman $x_p$   &    46  &  196  &  171  &  183  &  239  &  211  &  370  &   365  \\ 
$\pto$          &    19  & 1067  & 1010  & 1023  &  864  &  730  &  110  &   116  \\ 
$\pti$          &    25  &  107  &   63  &   70  &  156  &   73  &  171  &   164  \\  \hline 
   sum          &   168  & 1592  & 1488  & 1563  & 1409  & 1190  & 1374  &  1476  \\  
with cuts       &   137  &  355  &  304  &  382  &  372  &  299  &  -    &   -    \\  \hline             
\end{tabular} 
\end{center} \end{table} 

The large overall $\chi^2$ values indicate that, despite parameter tuning, it is very 
difficult for the models to reproduce the data with a 1 to 2\% accuracy. The JETSET and ARIADNE
models underestimate the $\pto$ tail by up to 30\% with deviations typically around
10 $\sigma$. This distribution is much better described by the HERWIG model.
On the other hand, the distributions of sphericity and thrust, 
and of the scaled momenta, $x_p$, of charged particles are better described by JETSET or 
ARIADNE. HERWIG shows a significant excess (10 $\sigma$) of low-thrust events and does not 
adequately describe the production of heavier particles like K-mesons and baryons.     

The total uncertainties of the parameters in Tables \ref{JStune}--\ref{HWtune} are determined as 
follows. First, the systematic uncertainties of the fitted distributions are included in the
$\chi^2$ minimization and thus propagated into an uncertainty on the parameters. 
Second, the fit procedure is changed by varying the fit regions and, in the case of HERWIG,
the set of baryons included in the fit. The largest changes in the fitted parameters are added 
quadratically to the first contributions.    

In order to test whether or not colour reconnection improves the overall description,
the CR parameter of each model is fitted simultaneously with the other fragmentation parameters.
The fit of the JETSET+GAL model has a $\chi^2$ minimum with respect to the CR parameter 
$R_0$ at approximately 0.04. If this parameter is fixed at the recommended value of 0.10, 
the changes in the other parameters are mainly a larger value for the parton shower cut-off $Q_0$ 
and a smaller value for the $b$ parameter.                    
The optimal value for the CR parameter $N_{\mathrm{C}}^2$ in the AR1 model is $8.4\pm1.1$, 
consistent with the default value of 9. 
Although the changes in total $\chi^2$ are not large, the distributions of global variables show 
a slight preference for the string-based colour reconnection models. For the HERWIG-CR model, 
the $\chi^2$ increases continuously with increasing $\Preco$ parameter. However,   
the global variables like event-shape and particle momentum distributions considered here may 
not be specific enough for testing CR models.

\section{Effects of colour reconnection in Z $\rightarrow $ hadrons} 
According to the string-based models AR1 and GAL, colour reconnection leads to shorter strings and 
thus to less particle multiplicity, on average. The changes of particle multiplicity and of the 
particle momentum distributions can be compensated, at least partially, by re-tuning the important   
fragmentation parameters. Clearly, more specific variables are needed to test CR models.  

Studies of AR1 and GAL at the generator level show that the fraction of events with one  
or more reconnections, f(reco), strongly rises with the number of partons and with the number of 
hadronic jets in the event. 
Also, a minimum of four partons in the final state are needed to perform a reconnection.  
This suggests that CR is related to the presence of gluon jets. The simplest configuration
which is expected to be favourable for testing CR models is therefore a three-jet event with one
of the jets being a well separated and energetic gluon jet. 

It is therefore important to check how well the gluon jet is described by the QCD models. 
The properties of gluon jets in comparison to those of quark jets have been studied in great 
detail at LEP \cite{ydis_OPAL,sj_ALE,qgprop_ALE,sv_DEL,sv_OPAL}. These studies include the 
fragmentation function and its scale dependence, the distributions of particle multiplicity and 
rapidity and the sub-jet structure. In general, good agreement between the data and the QCD 
model predictions is found. An exception is the fragmentation function at large $x$ where the 
predictions are systematically low \cite{sj_ALE,sv_OPAL}. 

A very specific and rare class of three-jet events in which the gluon jet exhibits a rapidity gap   
and in which the hadronic system beyond the gap has zero electric charge,  
has been proposed by the OPAL collaboration \cite{ygap_OPAL}, referring to theoretical ideas of
Ref.\ \cite{CR2}, as a signature for possible effects of colour octet neutralisation 
or colour reconnection. If a reconnection occurs according to the GAL or AR1 
models, a gluon jet, which in general consists of several gluons, often hadronizes as a closed 
string separated from the $\qqbar$ string by a rapidity gap. The signature is an increased rate 
of gluon jets exhibiting a rapidity gap and zero electric charge of the system beyond the gap.

\section{ALEPH detector, data and Monte Carlo samples}
A description of the detector and its performance can be found elsewhere \cite{AL_det,AL_perf}. 
Charged particles are detected in the central part consisting of a precision silicon strip 
detector (VDET), a cylindrical drift chamber (ITC) and a large time projection chamber (TPC). 
Jets originating from heavy quarks, in particular b-quarks, are identified with a lifetime
tagging algorithm which takes advantage of the 3-dimensional impact parameter resolution of
charged particle tracks. 
The tracking chambers are surrounded by the electromagnetic calorimeter (ECAL) located  
inside the magnet coil, and the hadronic calorimeter (HCAL). 

The information from the tracking detectors and the calorimeters is combined in an energy-flow
algorithm \cite{AL_perf}. 
This algorithm provides, for each event, a list of reconstructed objects, classified 
as charged particles, photons and neutral hadrons, and called {\it energy flow objects} in the 
following. Objects reconstructed in the luminosity detectors are omitted.     

Multihadronic Z decays are preselected by requiring at least five good charged tracks whose 
energy sum exceeds 10\% of the center-of-mass energy, $\Ecm$. 
Good tracks are defined as originating close to the 
interaction point (with transverse impact parameter $|d_0|<2$ cm and longitudinal impact 
parameter $|z_0|<10$ cm), having at least four TPC hits and a polar angle such that 
$|\cos\theta|<0.95$.  Residual backgrounds from $\tau$ pair and $\gamma\gamma$
events are reduced to a negligible level by requiring the events to have at least 14 energy 
flow objects whose energy sum, $\Evis$, exceeds 50\% of $\Ecm$. 
This selection yields 3 378 000 hadronic events from the ALEPH data collected at the Z peak 
($\Ecm \approx 91.2$ GeV) during the years 1992-1995.   

The analysis of the present paper relies on comparisons of data with QCD model calculations. 
Monte Carlo events were generated using tuned parameters as given in section 3.  
These events were passed through a full simulation of the ALEPH detector and  
were subject to the same reconstruction and analysis programs as the data. The numbers of 
hadronic Z decays generated for each of the models are given in Table \ref{mcstat}. 
If in the following the GAL model is mentioned, the version with $R_0=0.10$ is implied unless  
stated otherwise. 
 
\begin{table}  \begin{center}  
\caption{\label{mcstat} \small Statistics generated for each QCD Monte Carlo model.}  \vspace{1ex}
\begin{tabular}{|l|l|}   \hline
  QCD model            &  million events  \\  \hline 
JETSET                 &  7.6   \\
JETSET+GAL, $R_0=0.04$ &  0.5  \\  
JETSET+GAL, $R_0=0.10$ &  3.4  \\
JETSET+BE$_{32}$       &  0.5  \\ 
ARIADNE AR0            &  3.4  \\
ARIADNE AR1            &  3.4  \\
HERWIG                 &  0.5  \\
HERWIG CR              &  0.5  \\  \hline                  
\end{tabular} 
\end{center} \end{table}

\section{Analysis of energy-ordered jets}
\subsection{Three-jet event selection}
The Durham (or $\ksubT$) cluster algorithm \cite{Durh} is applied 
to the energy flow objects in order  
to determine the number of jets. The distance measure between any two particles is defined as 

\[  y_{ik}= 2 \min(E_i^2,E_k^2) (1-\cos\theta_{ik}) /\Evis^2 = (2 \ksubT/\Evis)^2.   \]   
A value of 0.02 has been chosen for the resolution parameter $\ycut$ as a compromise between well
separated jets and sufficient statistics. This value corresponds to $k_{\mathrm{T,cut}}=6.5$  
GeV for the case $\Evis=\Ecm$. This analysis deals with events which have exactly three jets. 
Their fraction is 23\% in data before any further cuts. 
Three-jet events in which more than 95\% of any jet energy is carried by photons are assumed 
to be hard bremsstrahlung off quarks and are rejected (0.7\% of the three-jet events).
To improve the particle acceptance, the event is only kept if the angle $\theta_j$ of each jet  
with respect to the beam direction satisfies $|\cos\theta_j|<0.90$.     

The jet energy resolution is improved by the following procedure.
The jets are first ordered according to their observed energies, $E_1>E_2>E_3$.
As the measured jet vectors in general do not exactly form a plane, an average
plane is constructed by taking 
the vector $(\vec{p}_1 \times \vec{p}_3)+(\vec{p}_3 \times \vec{p}_2)$ as
the normal to this plane. Jets 1 and 2 are projected into this plane.  
The jet energies are recomputed from the relative angles 
assuming energy-momentum conservation for massless jets. The jets are then ordered again,  
but this time according to the computed jet energies such that $E_1>E_2>E_3$. As further 
kinematic cuts the smallest of the interjet angles is required to be greater than 40 degrees and 
the smallest of the scaled jet energies, $x_j=2E_j/\Ecm$, to be greater than 0.1. These cuts
only become important for resolution parameter values smaller than 0.02 used for systematic  
checks. All these criteria result in \mbox{539 000} three-jet events ($=\Ntj$). 
Some of the jet properties are listed in Table \ref{eojets}.    

The information on the shower development of JETSET Monte Carlo events is used to estimate the 
probability, $\Pg$, of a jet to be the gluon jet.
The primary quarks from Z $\rightarrow \qqbar $ (after termination of the parton shower)
are assigned to the reconstructed jets by means of the smallest angle. 
The jet which has no quark assigned to it is then called the gluon jet.  
The least energetic jet (jet 3) has an average computed energy of 17.7 GeV.   
It represents a gluon jet in about 69\% of the cases. The highest energy jet (jet 1) is dominantly 
a quark jet with flavour composition given by the electroweak Z couplings. 
The selected sample comprises a wide range of kinematic configurations.  

\begin{table}[h]  \begin{center}  
\caption{\label{eojets} \small Some properties of energy-ordered jets in data.}   \vspace{1ex} 
\begin{tabular}{|c|cc|c|c|}   \hline
 jet number & $\langle E_{\mathrm{jet}} \rangle$, GeV & 
                                  r.m.s.   & $\nch$       &   $\Pg$     \\  \hline
  1         &  40.8            &  2.7      &  8.81        &   0.059     \\
  2         &  32.7            &  4.5      &  8.28        &   0.248     \\
  3         &  17.7            &  5.1      &  7.02        &   0.693     \\  \hline     
\end{tabular}      
\end{center} \end{table}    

\subsection{Jet charge distributions}
The jet charge is defined here as the sum of the charges $q_i$ (in units of the elementary charge) 
of the particles which are assigned to jet {\it j} by the cluster algorithm:  

\begin{equation}  Q_{j}=\sum_{i=1}^n q_i =n_+ - n_- \end{equation}    
with $n$ being the number of charged particles in the jet under consideration and $n_+(n_-)$ the
number of particles with charge +1 (--1). This definition does not explicitly depend on the
particle momenta. It has previously been used by the OPAL collaboration \cite{Qj_OPAL}
in a study of quark and gluon jet charges. 
  
The $Q_{j}$ distribution is influenced by many factors. The charge of the underlying hard 
parton is expected to be reflected in the charges of the produced hadrons. Therefore the charge 
of a gluon jet should be zero on average. This is also true for quark-initiated jets because 
quarks are not distinguished from antiquarks in this analysis, which enter with equal frequency. 
An important property is the r.m.s. width of the distribution which depends on
the mean particle multiplicity, on the jet environment and on the jet finder used. 
In addition the width is limited by local charge compensation in jet 
fragmentation together with charge conservation in the whole event.  
Also, Bose--Einstein correlations have an influence on the width. 
Experimental effects in general tend to smear and to shift the distribution. 

\begin{figure}[b]  \begin{center}
\epsfig{file=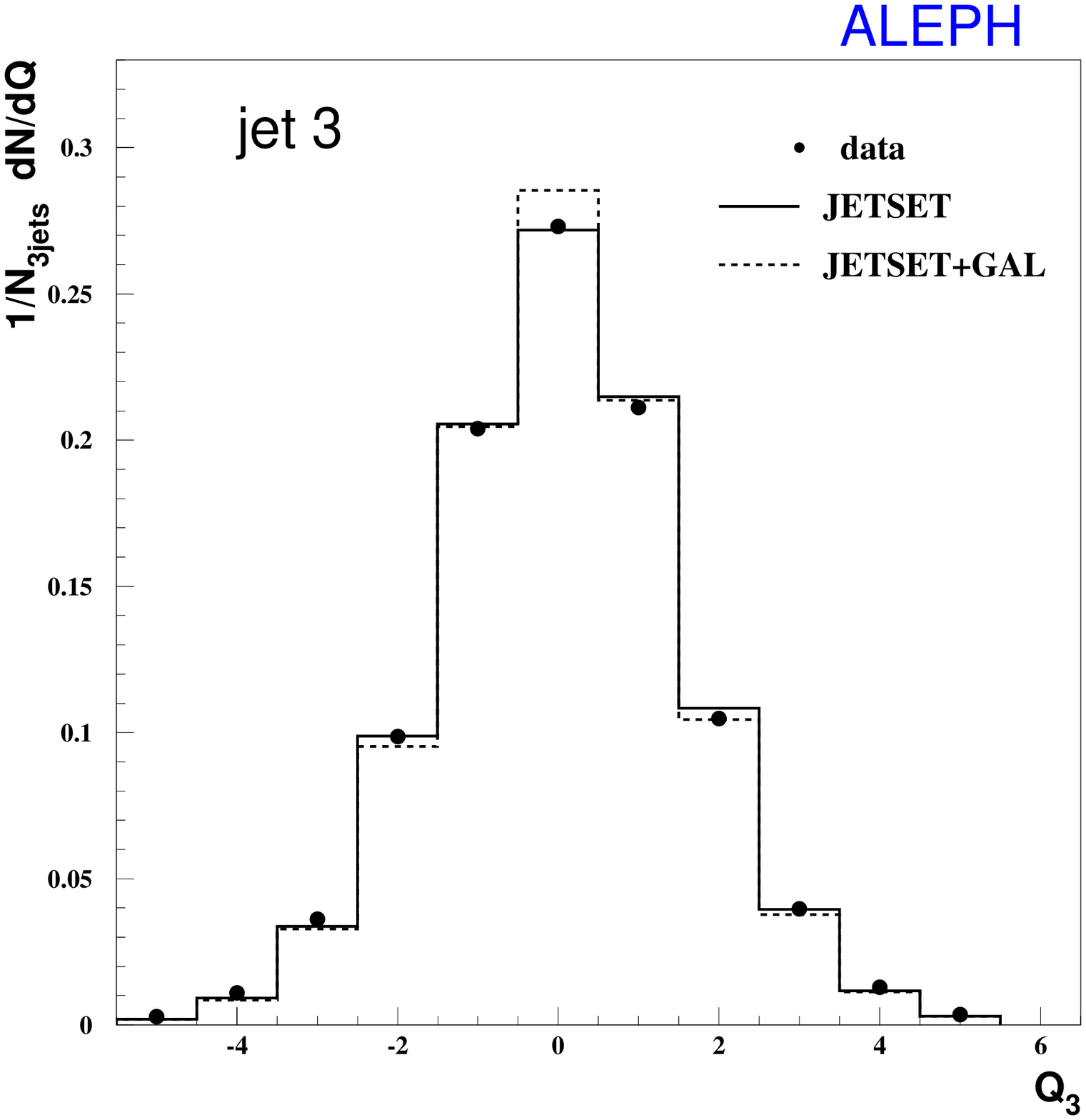,width=10cm}
\caption{\label{qj3}  \small The charge distribution of jet 3 (the lowest energy jet) compared to 
JETSET predictions without and with colour reconnection.}   
\end{center}  \end{figure}

The measured $Q_{3}$ distribution of jet 3 is shown in Fig.\ \ref{qj3} 
for the full data sample, together with the JETSET prediction.
The data are rather well described by the simulation, in particular the fraction of neutral 
($Q_{3}=0$) jets. The data distribution is not symmetric around 0, but slightly shifted 
towards positive values, a feature which is also reproduced by the simulation and which can be 
explained by protons from nuclear interactions in the detector material. The same is true for jets 
1 and 2.  The colour reconnection model GAL predicts a higher rate of neutral jets in the
gluon-enriched jet (jet 3) than is seen in the data.  

\subsection{Rapidity gaps}  
The particle distribution in jets is analysed in terms of rapidity 

\[  y=\ln \left( \frac{E+p_{\mathrm{L}}}{\sqrt{m^2+p_{\mathrm{T}}^2}} \right) .    \]  
The longitudinal and transverse momentum components ($p_{\mathrm{L}}$ and $p_{\mathrm{T}}$)
are measured with respect to the jet axis,
which is defined as the vector sum of all energy flow objects assigned to the jet. 
The pion mass is assumed for all charged particles except for those identified as electrons or
muons, where $m_{\mathrm{e}}$ or $m_{\mu}$ is used. Neutral particles are assumed to be massless.
They are included only if their energy exceeds 0.6 GeV because of the poor agreement between   
data and simulation at low energies.       

\begin{figure}[b]  \begin{center}
\epsfig{file=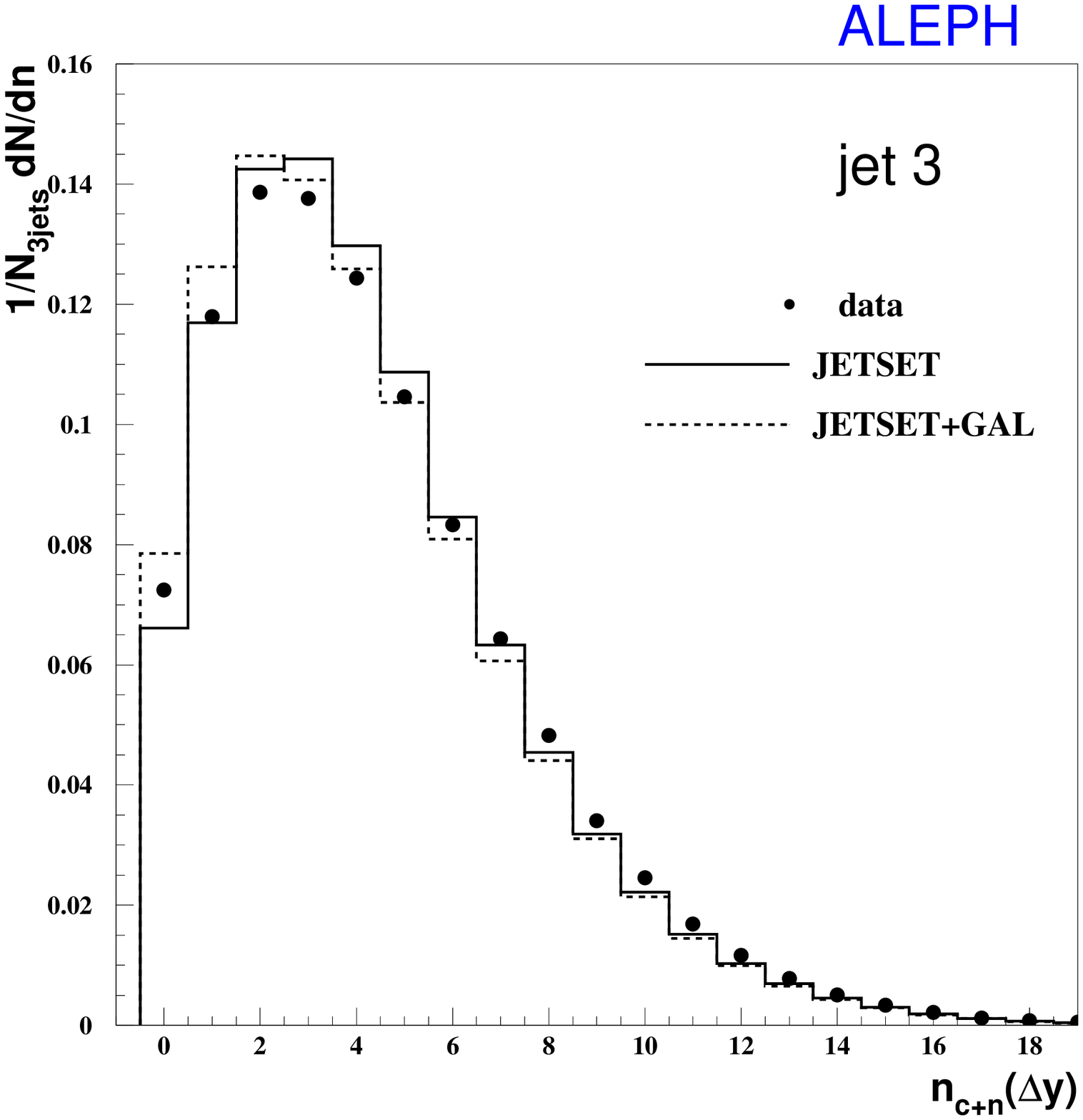,width=10cm}
\caption{\label{ncnj3_dy} \small The distribution of the number of charged plus neutral particles 
 in the  rapidity interval $0 \leq y \leq 1.5$ for jet 3,  
 compared to JETSET without and with CR.} 
\end{center}  \end{figure}

Figure \ref{ncnj3_dy} shows the multiplicity distribution of charged plus neutral particles, 
$n_{\mathrm{c+n}}(\Delta y)$, for jet 3 within a central rapidity interval 
$0 \leq y \leq \yupp$ with the upper limit chosen as $\yupp=1.5$.  
About 7\% of the jets in data have zero particles in this interval, as seen in the 
leftmost bin in the figure. 
Starting from $\yupp=1.0$, the rate of these `gap-jets' falls nearly exponentially with  
increasing interval size. The rate is seen to be sensitive to CR effects :   
JETSET predicts too few and GAL predicts too many of these jets. The rate of jets with a 
rapidity gap obviously depends on the shape of the multiplicity distribution. 
The r.m.s. width of the distribution is slightly underestimated by the QCD models.     

A second definition of a rapidity gap, based on the maximum rapidity distance, 
$\Delta y_{\mathrm{max}}$,
between adjacent particles in a jet has been proposed in  \cite{ygap_OPAL}. Since the system 
beyond the gap selected in this way is found to consist dominantly of one particle only and is found
to be less sensitive to CR effects, it is not investigated further.   

\subsection{Charge distributions of jets with a rapidity gap}
\begin{figure}[b]  \begin{center}
\epsfig{file=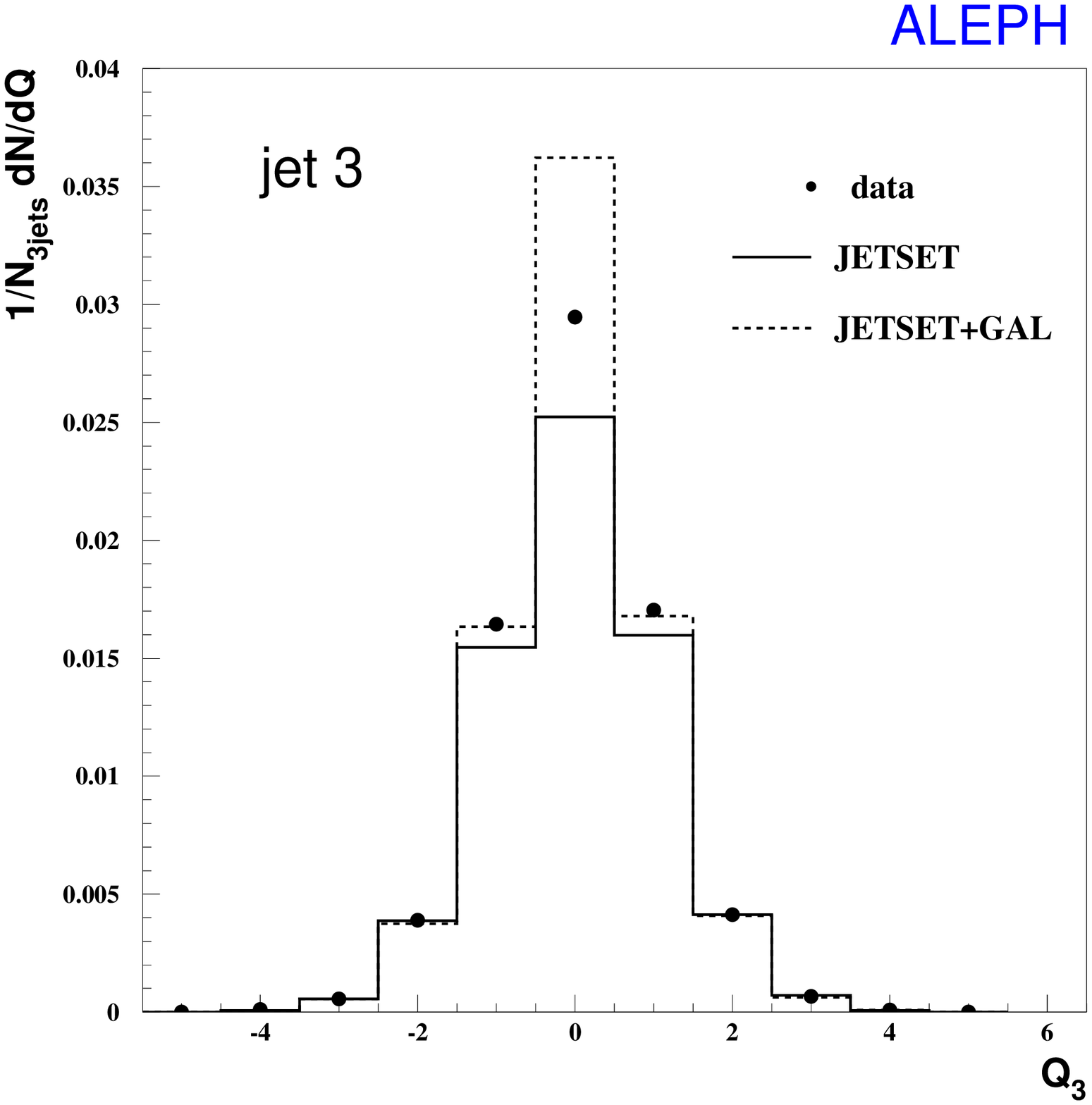,width=10cm}
\caption{\label{qj3_dycngap} \small The charge distribution of jet 3 for a rapidity gap in
 $0 \leq y \leq 1.5$.}

\epsfig{file=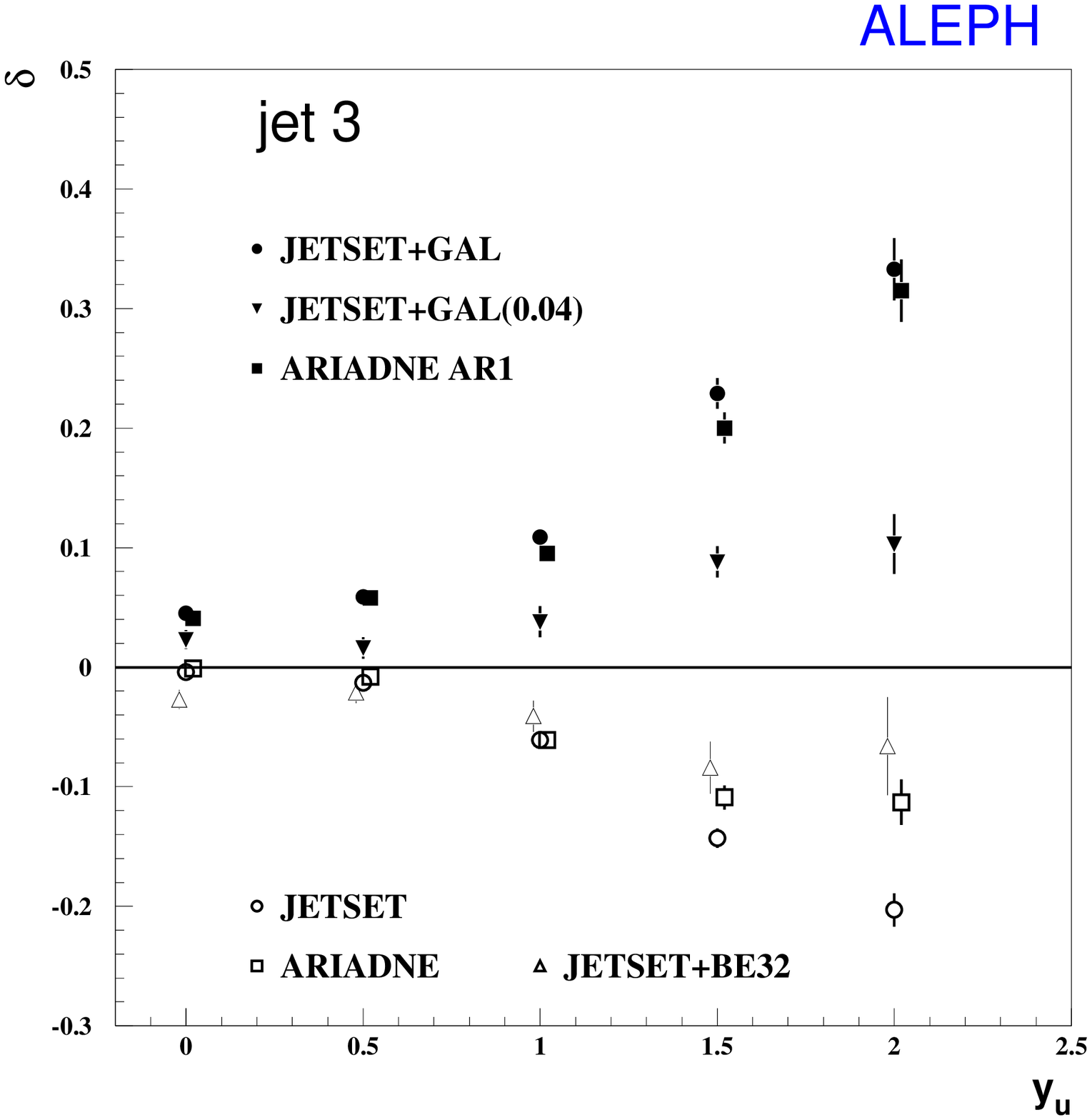,width=10cm}
\caption{\label{delta_eo_j3} \small 
 The relative model--data differences $\delta$ in the rate $r(0)$ of neutral jets   
 for jet 3 as a function of the upper limit of the rapidity gap. The points at $y_u=0$ correspond  
 to the full sample. Statistical uncertainties of the data and the Monte Carlo calculations 
 are shown as error bars.}  
\end{center}  \end{figure}

The charge distribution of jets which exhibit a central rapidity gap as defined above 
is shown in Fig.\ \ref{qj3_dycngap} for jet 3. The jet charge is computed from charged particles  
with rapidities $y>\yupp$. The fraction of particles moving backwards ($y<0$) is negligibly
small (of order $10^{-4}$).  
The distribution is normalized to the total number of three-jet events, $\Ntj$,
which means that the rate of gap-jets also enters the comparison of data with Monte Carlo models.
The distribution in Fig.\ \ref{qj3_dycngap} is narrower than the corresponding one for the full sample 
(Fig.\ \ref{qj3}) due to the smaller average particle multiplicity.  
The rate of neutral jets with a rapidity gap (Fig.\ \ref{qj3_dycngap}) shows an even higher    
sensitivity to CR than the rate of gap-jets alone.  
None of the models agrees well with the data : while the GAL model predicts too many neutral jets,
standard JETSET predicts too few of them.    \vspace{1ex}  

\begin{figure}[b]  \begin{center}
\epsfig{file=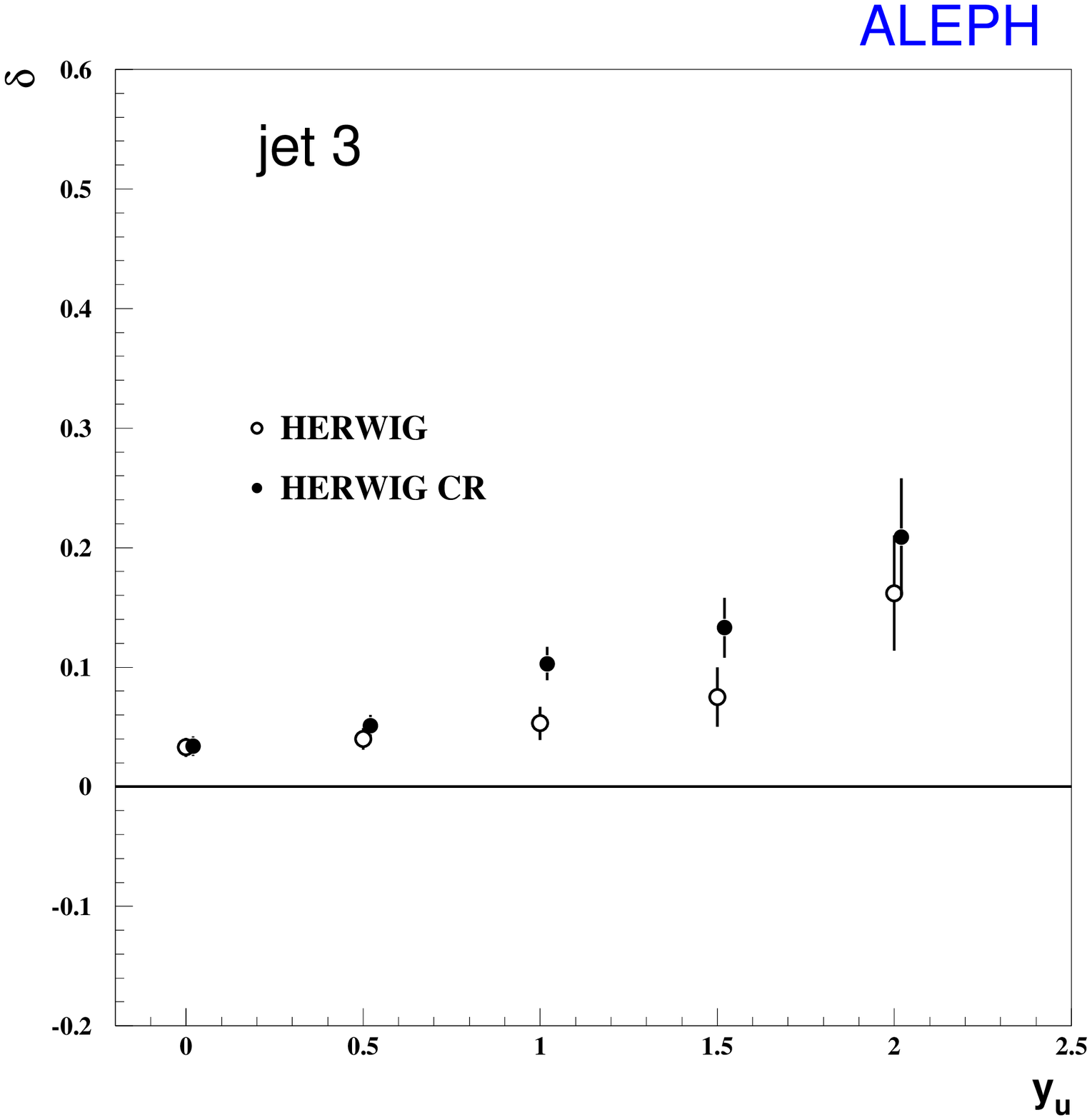,width=10cm}
\caption{\label{delta_hw_j3} \small 
 The relative HERWIG--data differences $\delta$ in the rate $r(0)$ of neutral jets   
 for jet 3 as a function of the upper limit of the rapidity gap.} 
 
\epsfig{file=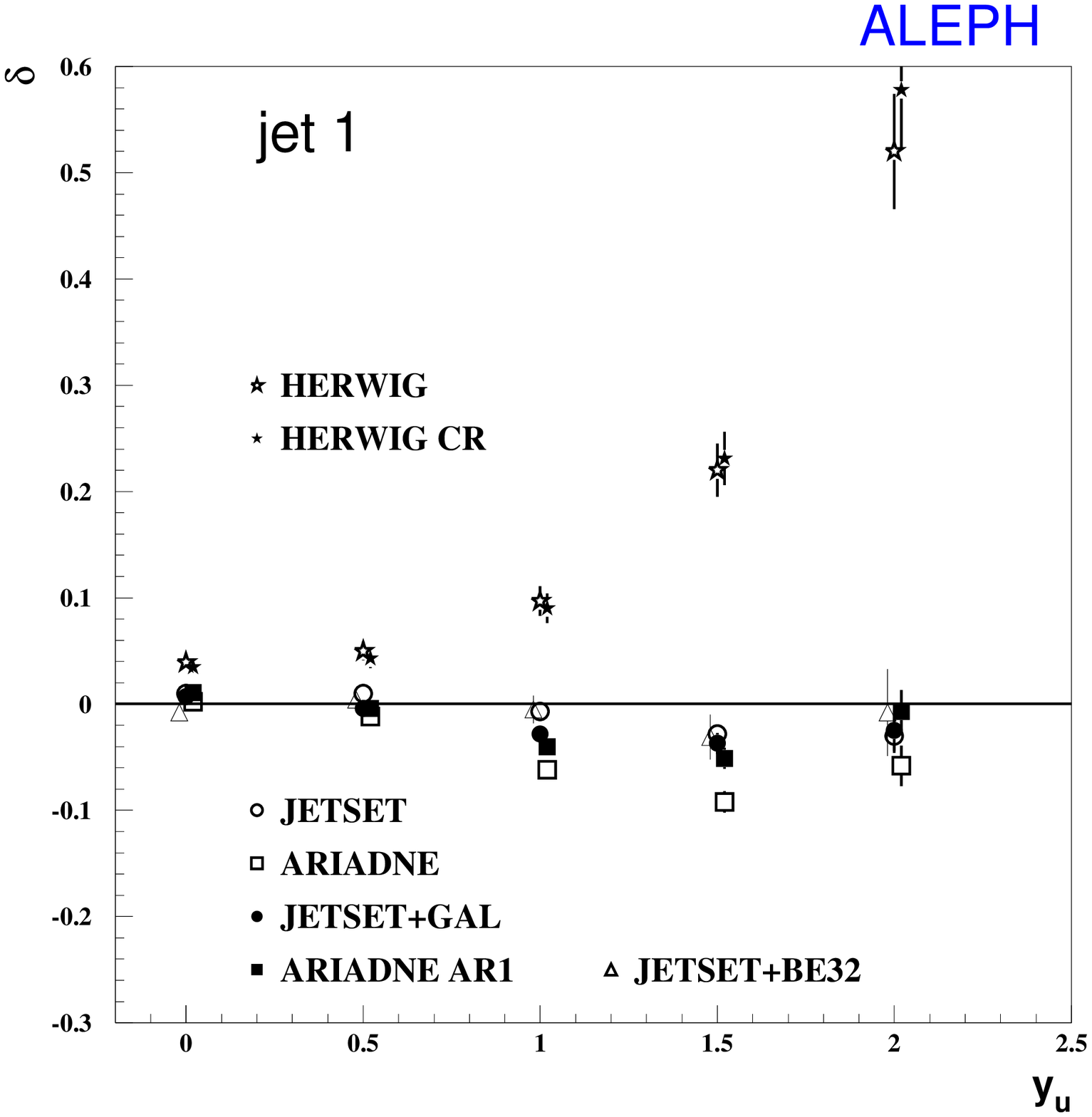,width=10cm}
\caption{\label{delta_eo_j1} \small
 The relative model--data differences $\delta$ in the rate $r(0)$ of neutral jets   
 for jet 1 as a function of the upper limit of the rapidity gap. The points at $y_u=0$ correspond  
 to the full sample. Statistical uncertainties of the data and the Monte Carlo 
 calculations are shown as error bars.}  
\end{center}  \end{figure}

The relative model--data differences of the rate $r(0)$ of neutral jets     
\begin{equation}  
  \delta = \frac{r(0)_{\mathrm{MC}}-r(0)_{\mathrm{data}}}{r(0)_{\mathrm{data}}}, 
  \quad\mbox{where}\quad  r(0)=\frac{N(Q_{j}=0, \Delta y \:\mbox{gap})}{\Ntj}    
\end{equation}
are shown in Fig.\ \ref{delta_eo_j3} for jet 3 as a function of the rapidity gap size.  
Note that the points are not statistically independent : each point represents a sub-sample of 
the previous point(s). Both the colour-reconnected and the standard versions of 
JETSET and ARIADNE are seen to diverge from the data with increasing gap size. 
For the case $\yupp=1.5$, the GAL model deviates from the data by 17 $\sigma$ (statistical). 
The CR effect in the AR1 model is slightly weaker than that in the GAL model.  
Colour reconnection as implemented in these two models and assuming default values for the 
strength parameters, is therefore disfavoured by the data. 
The GAL model with a smaller value for the strength parameter, $R_0=0.04$, still gives somewhat 
too high predictions. 
On the other hand, the rates predicted by standard JETSET and ARIADNE (without CR) 
are systematically low if there is a rapidity gap. The quantity $r(0)$ has little or no 
sensitivity to the HERWIG-CR model, as can be seen from Fig.\ \ref{delta_hw_j3}.  

The $\delta$ values for the quark-enriched jet (jet 1) are shown in Fig.\ \ref{delta_eo_j1} 
for comparison. As expected, this jet shows very little sensitivity to CR for all models. 
The data are quite well described by JETSET. 
The ARIADNE model shows some deviation in the case of a gap.   
HERWIG is unable to describe the data. This is related to the observation that HERWIG's 
particle rapidity distributions are systematically lower over a wide range of rapidity
and that the $Q_{j}$ distributions are narrower than the corresponding data distributions.  

The effect of Bose-Einstein correlations, as simulated with the tuned BE$_{32}$ model,
is to slightly broaden the charge distributions, leading to a downward shift of the $\delta$ 
values by about --0.02 for the full samples of jets 1 and 3. For events with a gap in jet 3  
(Fig.\ \ref{delta_eo_j3}) the shift in $\delta$ goes into the other direction.         

The results are less dependent on model-data discrepancies in the multiplicity distributions
if the $Q_j$ distributions of jets with a rapidity gap are normalized to unit area for both data 
and Monte Carlo models. The quantity studied is the fraction of neutral jets  
\begin{equation}
  f(0)=\frac{N(Q_{j}=0, \: \Delta y \:\mbox{gap})}{N(\Delta y \:\mbox{gap})},   
\end{equation}     
where $N(\Delta y \:\mbox{gap})$ is the number of jets with a rapidity gap. The model-data
differences are similar to those discussed above but statistically less significant. 
The predictions of the non-CR models JETSET and ARIADNE for the third jet are low also in this 
case, with a relative deviation from the data by about --6\% for a rapidity gap in   
$0 \leq y \leq 1.5$, confirming the results of a similar analysis by DELPHI \cite{ql_DELPHI}. 

\subsection{Corrected rate of neutral jets} 
The results presented so far indicate that none of the QCD models provides a satisfactory 
description of the data for jet 3 with a rapidity gap. If the data are to be described
by a colour reconnection model, the question then arises which value of the CR parameter 
is required by the data.
For this study, the parameter $R_0$ of the GAL model is considered, (Eq.\ (1)). 
In order to save computing time, the data are first corrected for the effects 
of the detector, of the reconstruction chain and the analysis cuts by means of a correction 
factor ($C$) derived from the JETSET simulation : 

\begin{equation}    r(0)_{\mathrm{data}}^{\mathrm{corr}}= r(0)_{\mathrm{data}} 
 \cdot \frac{r(0)_{\mathrm{MC,gen}}}{r(0)_{\mathrm{MC,sim}}} = r(0)_{\mathrm{data}} \cdot C,  
\end{equation} 
and similarly for $f(0)$.  
The calculation including full simulation of the detector is denoted as 'MC,sim'. The
calculation at the particle level of the event generator ('MC,gen') is done separately  
according to the following prescription.   
Hadronic Z events are generated without initial- and final-state photon radiation.
The Durham cluster algorithm is applied to all charged and neutral stable particles 
to select three-jet events. The jets are ordered according to their actual energies. Rapidities are
calculated using the actual particle masses. To define a rapidity gap, neutral particles with 
energies smaller than 0.6 GeV are omitted.   

\begin{table}[h]  \begin{center}  
\caption{\label{rcorr} \small Corrected data values for the rates of neutral jets in jet 3 with a 
 rapidity gap from 0 to 1.5. The first error is statistical, the second systematic.} \vspace{1ex} 
\begin{tabular}{|l|l|}   \hline   
          &    corrected  data   \\  \hline
$r(0)$    &    0.0234 $\pm$ 0.0004 $\pm$ 0.0015  \\
$f(0)$    &    0.437  $\pm$ 0.006  $\pm$ 0.010   \\  \hline
\end{tabular}      
\end{center} \end{table}  

The results for the lowest energy jet (jet 3) and requiring a gap in the rapidity range  
\mbox{0--1.5} are given in Table \ref{rcorr}. 
The quoted systematic error arises from model dependence and is estimated by taking 
the largest difference of the results when correcting with JETSET, GAL(0.04) or ARIADNE. 
These numbers can then be compared to those obtained from generator level calculations of the 
JETSET+GAL model for different values of $R_0$. The complication here is that the CR effect is not 
a one-parameter problem : the fraction of reconnected events depends strongly on the parton 
shower cut-off, $Q_0$, but only very weakly on the fragmentation parameter $b$. To ensure 
agreement of the average particle multiplicity with the data, at each $R_0$ value the parameters 
$\Lambda, \: \sigma$ and $b$ are re-tuned to global quantities, while keeping $Q_0$ fixed 
at 1.5 GeV.     

The comparison yields an optimal value for the colour reconnection parameter $R_0$ of about 0.02 
if the quantity $r(0)$ is used, and about 0.01 if $f(0)$ is used.   

\section{b-tag analysis}
A much higher gluon jet purity can be achieved by identifying both the quark and the antiquark   
jets in events of the type Z$\rightarrow \mathrm{b} \bar{\mathrm{b}} \mathrm{g}$,  
by utilizing the long lifetime of B-hadrons \cite{btag}. Based on the 
three-dimensional impact parameter and its significance measured for each charged particle track
the probability $\Pjet$ is computed that all charged tracks of a jet arise from the primary 
interaction vertex.  
The method of gluon-tagging is taken from  \cite{sj_ALE}. Three-jet events are selected in which 
two of the jets exhibit significant lifetime $(\Pjet<0.01)$, whereas the remaining jet does not 
$(\Pjet>0.01)$. This latter jet then represents the tagged gluon jet. No energy ordering is 
applied here.   
 
The event selection and the computation of the jet energies are the same as already described 
in Sect.\ 6.1 except that all three jets are required to fall in the geometrical
acceptance of the VDET $(|\cos\theta_j|<0.766)$ and that the scaled gluon jet energies, 
$x_{\mathrm{g}}$, are restricted to values smaller than 0.85. 

The fraction of three-jet events surviving the lifetime tag condition alone is 7.6\% in data,  
to be compared to 7.4\% in the JETSET simulation. About 24600 tagged three-jet 
events are selected from data. According to JETSET Monte Carlo calculations 95\% of
the events actually arise from primary b-quark pairs, 4.2\% from c-quark pairs and the rest from  
light quark pairs.  
The tagged gluon jet energies are distributed with mean value 19.8 GeV and an r.m.s. of 7.8 GeV. 
The tagged gluon jet is estimated to be the `true' gluon jet in 97.2\% of the cases on average. 
Since the gluon purity is found to drop rapidly at high energies,
the scaled jet energies are restricted to values $x_{\mathrm{g}} \leq 0.85$ as mentioned above.
A sub-sample of 18700 tagged gluon jets are lowest energy jets.
This sub-sample is therefore in common with the jet-3 sample of the energy-ordered  
analysis presented in Sect.\ 6.      
 
\begin{figure}[b]  \begin{center}
\epsfig{file=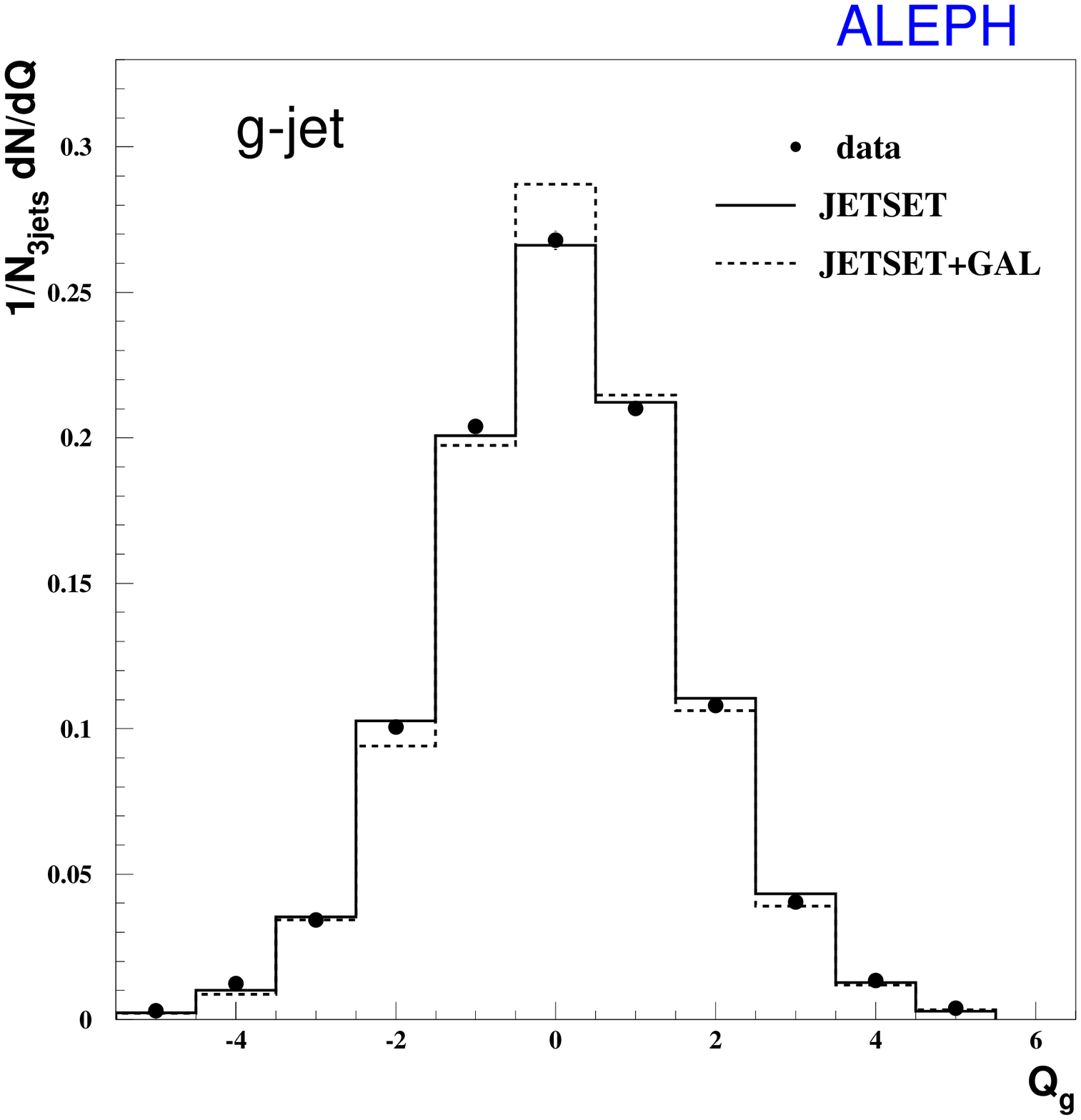,width=10cm}
\caption{\label{qjg}  \small The charge distribution of the anti-b tagged gluon jet compared to 
JETSET predictions without and with colour reconnection.}   
  
\epsfig{file=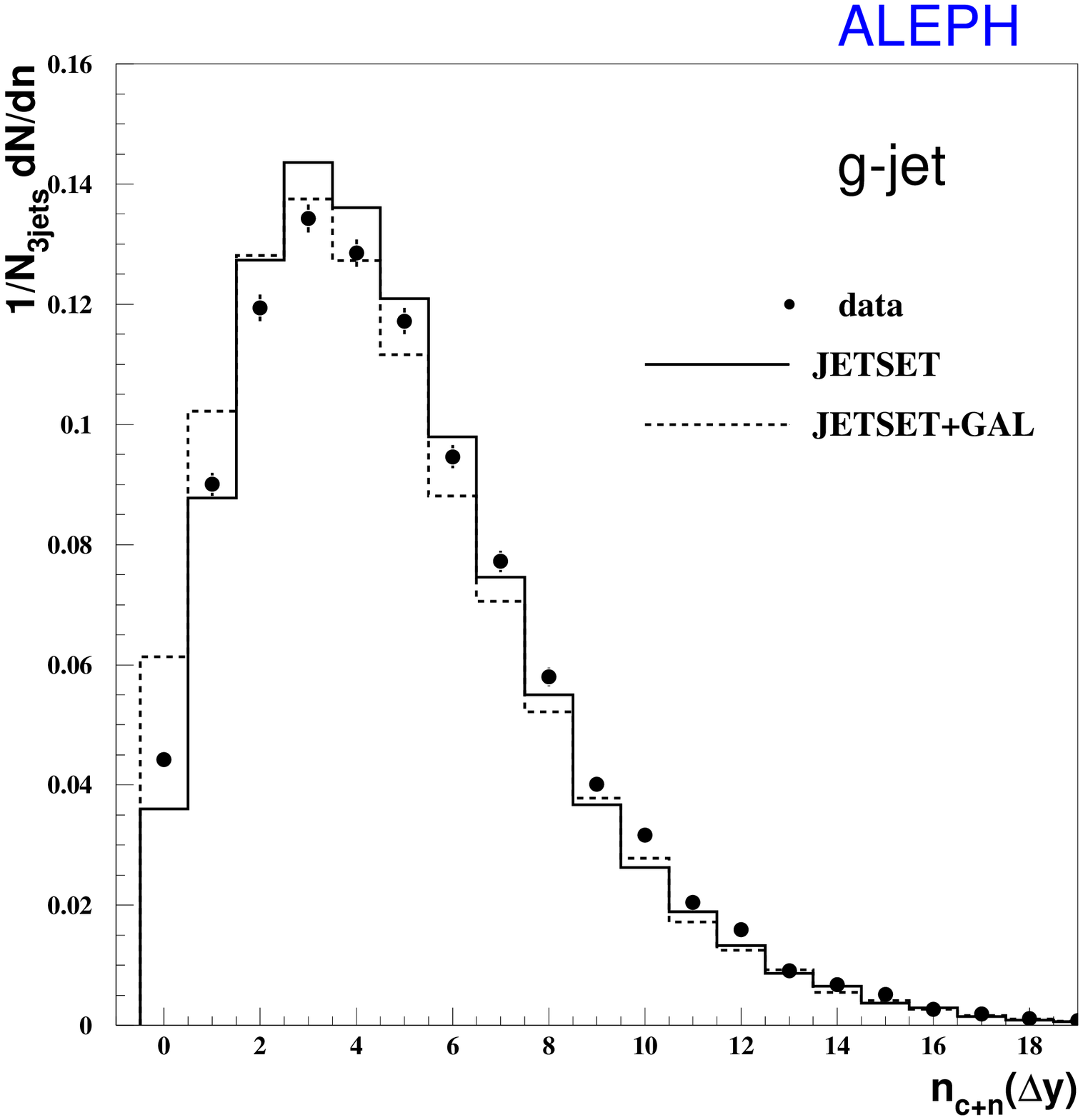,width=10cm}
\caption{\label{ncnjg_dy} \small The distribution of the number of charged plus neutral particles in  
 the rapidity interval $0 \leq y \leq 1.5$ for the gluon jet, 
 compared to JETSET without and with CR.}  
\end{center}  \end{figure}

In order to test colour reconnection models the same analysis is performed on the tagged gluon jet 
as on the least energetic jet in Sect.\ 6. Due to the higher gluon purity the results
are expected to be less dependent on the quark jet background in the sample. Figure \ref{qjg} shows  
the charge distribution of the tagged gluon jets. The rate of neutral ($Q_{\mathrm{g}}=0$) jets 
is well described by JETSET. The CR model predicts a clear excess with respect to the data.
Figure \ref{ncnjg_dy} shows the multiplicity distribution of charged plus neutral particles within  
the central rapidity interval $0\leq y\leq 1.5$. The rate of jets having zero particles in this 
interval (rapidity gap) shows sensitivity to CR. 

\begin{figure}[b]  \begin{center}
\epsfig{file=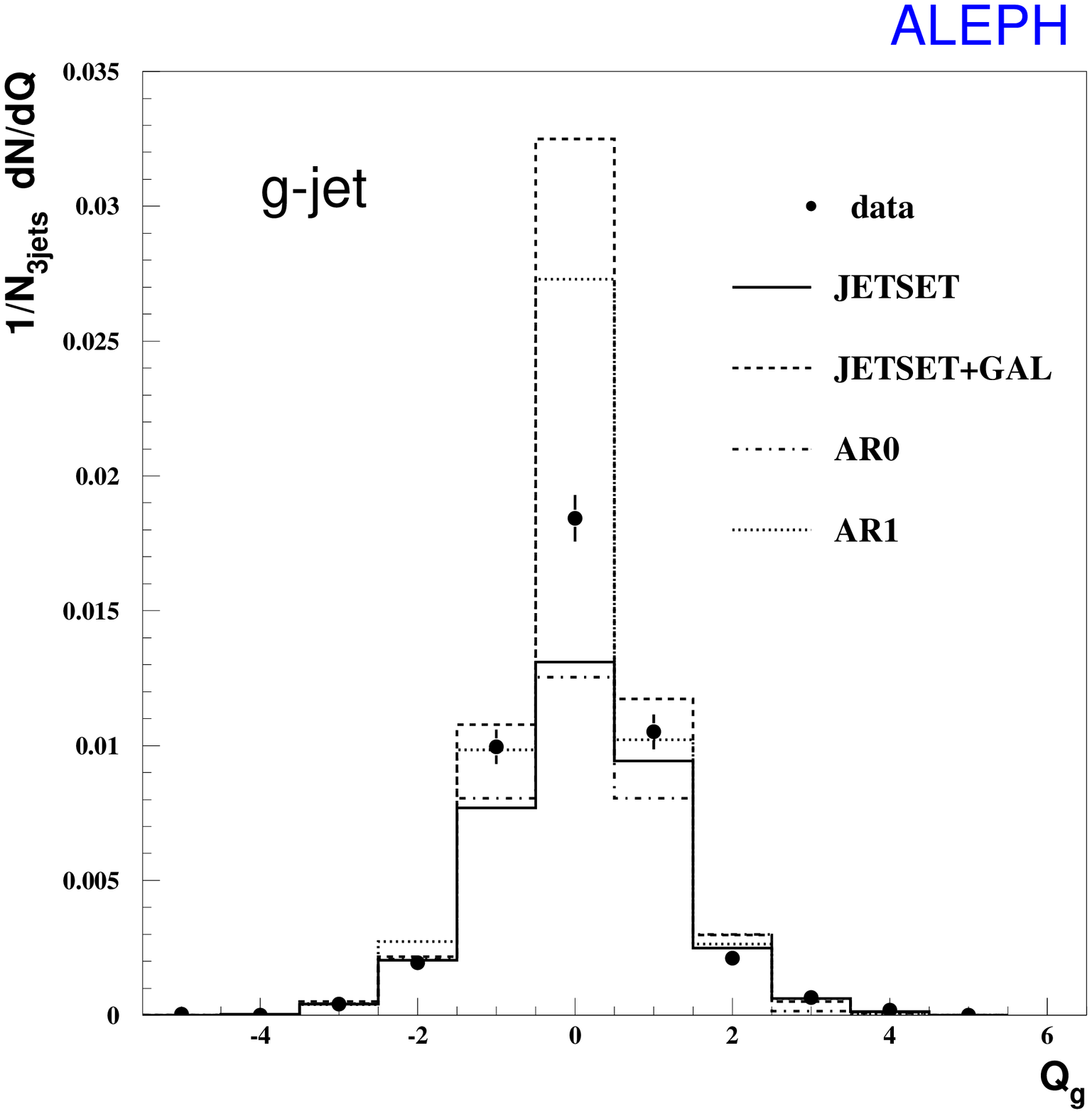,width=10cm}
\caption{\label{qjg_dycngap}  \small The charge distribution of gluon jets with a 
 rapidity gap in $0 \leq y \leq 1.5$.} 

\epsfig{file=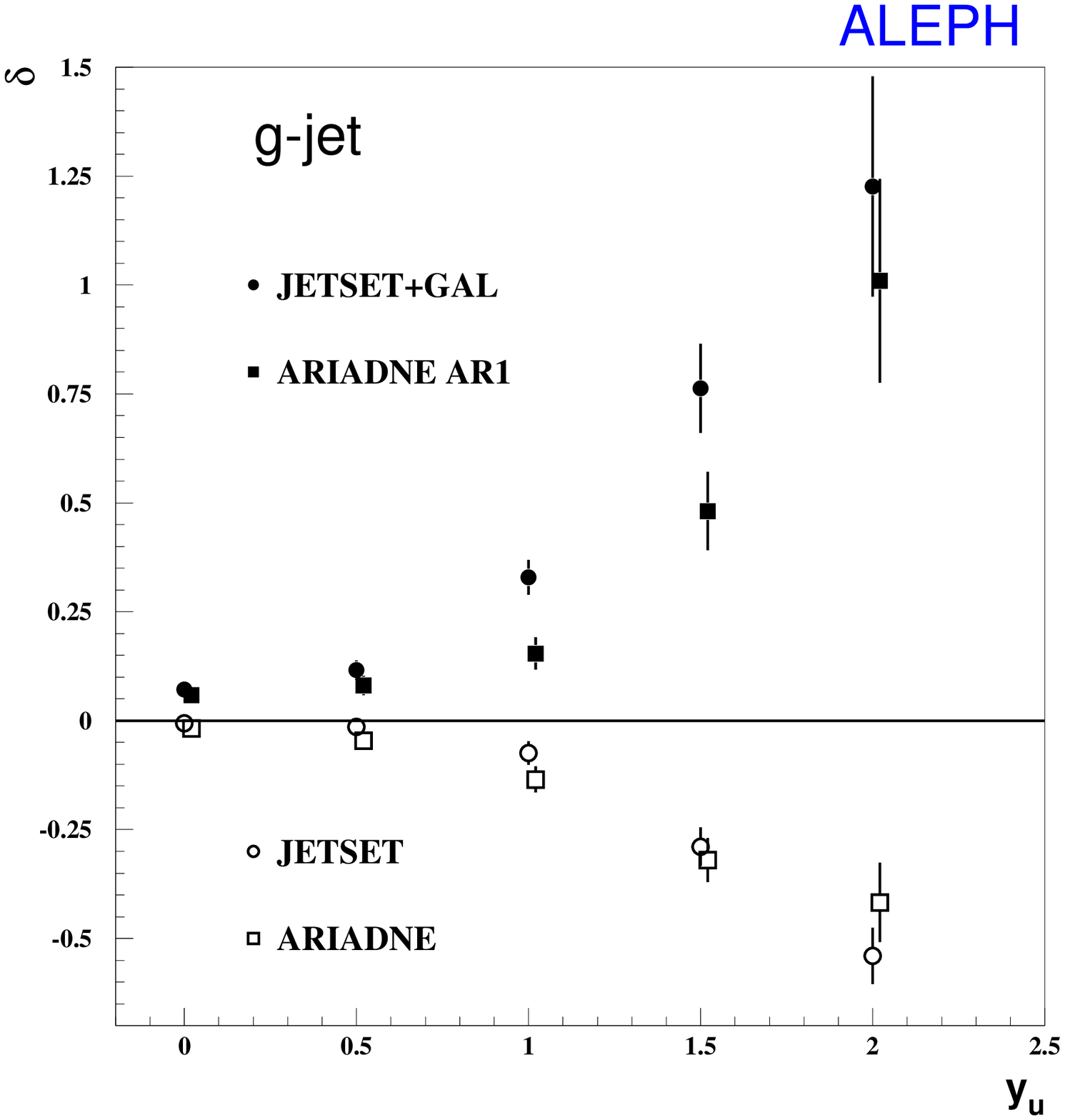,width=10.0cm}
\caption{\label{delta_bt_jg} \small 
 The relative model--data differences $\delta$ in the rate $r(0)$ of neutral  
 jets as a function of the upper limit of the rapidity gap for the gluon jet sample.  
 The points at $y_u=0$ correspond to the full sample. Statistical uncertainties of the data
 and the Monte Carlo calculations are shown as error bars.}    
\end{center}  \end{figure}

The charge distribution of the jets exhibiting a rapidity gap is shown in Fig.\ \ref{qjg_dycngap}.
This distribution is normalized to the number of tagged three-jet events and contains 1090 jets in 
data. Clearly, the CR models GAL and AR1 predict too many neutral gluon jets. 
The relative model--data differences $\delta$ (Fig.\ \ref{delta_bt_jg}) are numerically larger
than those from the energy-ordered analysis and reach values around 1.
The results of the b-tag analysis confirm those of the energy-ordered analysis although with less   
statistical significance.  
At $\yupp=1.5$ the GAL model deviates from the data by about 7 $\sigma$ (statistical).   

\begin{figure}[b]  \begin{center} 
\epsfig{file=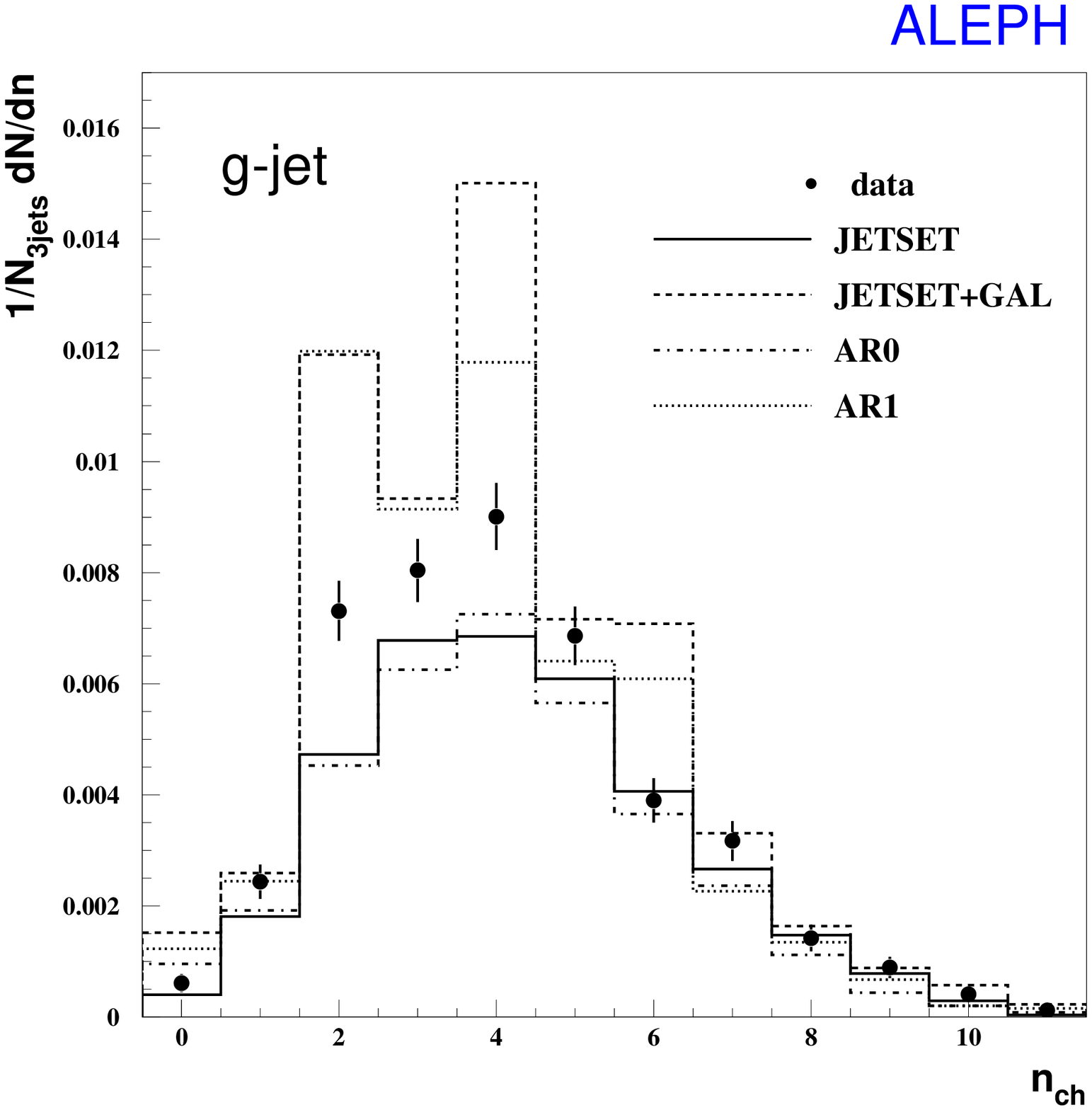,width=10cm}
\caption{\label{nchjg_dycngap}  \small The charged multiplicity distribution of gluon jets  
 with a rapidity gap in $0 \leq y \leq 1.5$.} 
\end{center}  \end{figure}

The distribution of the charged track multiplicity in gluon jets with a rapidity gap is shown in  
Fig.\ \ref{nchjg_dycngap}. The data do not exhibit clear evidence for spikes at the even values  
$n_{\mathrm{ch}}=2$, 4 and 6 as predicted by the CR models GAL and AR1.

\section{Systematic checks}
The energy-ordered analysis has been repeated by varying the definition of the jets and the definition
of the rapidity gap in the following ways :
\begin{itemize}
\item The Durham jet resolution parameter, $\ycut$, has been varied in the range from 0.01 to 0.03.
\item A different jet finder, the JADE algorithm, has been used with resolution parameter 
  $\ycut=0.08$, which leads to a similar three-jet event rate. 
\item A re-assignment of particles to jets was performed on the basis of the smallest angle with 
  respect to the jet axes. Afterwards the jet four-momenta are re-computed. This mainly affects 
  particles at large angles to the jet axes which are of importance for defining a rapidity gap. 
\item The rapidity gap has been defined with charged tracks only. 
\item The shifted rapidity regions (0.5--1.5) and (1.0--2.0) were also used to define the gap.   
\end{itemize}
In all five cases the qualitative features of the results remain the same, although the numerical 
values of the measured quantities may change.    

Possible inadequacies of the detector simulation have been studied by varying, in data and   
simulation, some of the cuts on the charged tracks 
and the cut on the polar angle of the jet axes according to Table \ref{cutvar}. 
\begin{table}  \begin{center}  
\caption{\label{cutvar} \small Experimental cuts varied for systematics studies.}  \vspace{1ex} 
\begin{tabular}{|c|cc|}   \hline
 cut        & standard value  & changed value  \\   \hline
 $|d_0|$, (cm)     &  2.0      &   0.5       \\
 $|z_0|$, (cm)     &  10.0     &   5.0       \\
 $N_{\mathrm{TPC}}$&  4        &    6        \\
 $|\cos\theta_j|$  &  0.90     &  0.80       \\   \hline
\end{tabular}      
\end{center} \end{table}    
This has been done only for the normalized jet charge distributions. In all cases,
the model--data differences only change within their statistical errors. 
 
The model--data comparison of the rate of jets with a rapidity gap relies on the calibration
of the average particle multiplicity at the detector level. 
In all hadronic events, the average charged particle 
multiplicity, $\nch$, of all generators, including HERWIG, agrees with the data to an 
accuracy below 1\% which, of course, is a result of the model parameter tuning. For JETSET 
and ARIADNE this also holds for three-jet events, and in addition for each of the three jets  
individually.   
In contrast, HERWIG exhibits a $1.5 - 2.5$ \% deficit in each of the three jets of three-jet  
events, thus giving less reliable predictions.

\newpage  
\section{Conclusions}
Three-jet final states from $\epem$ annihilation into hadrons are used to test QCD models of  
colour reconnection (CR). From a sample of approximately $3.4$ million hadronic Z decays,
collected by the ALEPH detector at LEP, three-jet events are selected by the $\ksubT$
cluster algorithm with a fixed cut-off, $\ycut=0.02$. The distributions of particle 
multiplicity in jets in fixed rapidity intervals and the distributions of the jet charge 
are investigated. The main analysis is based on jet energy ordering. According to JETSET, 
the least energetic jet (jet 3) represents a gluon jet in about \mbox{69 \%} of the cases.  

Gluon jets with a large gap in rapidity and with zero electric charge of the system beyond the gap 
constitute a particularly sensitive test. The rate, $r(0)$, of these jets is measured as a  
function of the rapidity gap size and is compared to predictions of globally tuned QCD generators.   
The rate $r(0)$ is predicted too high by the string-based CR models JETSET+GAL and ARIADNE AR1
when using default values for the CR strength parameters.  
Thus these two models are disfavoured by the data as already observed in Ref.\ \cite{ygap_OPAL}. 
On the other hand, the rate is too low for the standard, non-CR versions of these generators, 
in agreement with a similar analysis in Ref.\ \cite{ql_DELPHI}.   
The $R_0$ parameter of the GAL model is constrained by the data to the range $0.01-0.02$.  
The quantity $r(0)$ turns out to be insensitive to CR as implemented in the HERWIG-CR model.   

In a separate analysis of the same data, two b-quark jets are positively identified using 
lifetime tagging, thus achieving a very high gluon jet purity of about 97 \% for the remaining jet. 
The results confirm those of the energy-ordered analysis though with less statistical significance.    

Assuming that the physics of colour rearrangements is similar in hadronic Z decays and in WW 
decays, this result would imply that the CR models as implemented in JETSET+GAL and ARIADNE 
overestimate the systematic shift of the measured W boson mass. 

\section{Acknowledgements}
We wish to thank our colleagues from the CERN accelerator divisions for the successful operation 
of LEP. We thank L. L\"onnblad, J. Rathsman, T. Sj\"ostrand and B.R. Webber for discussions 
on the theoretical aspects.
We are indebted to the engineers and technicians in all our institutions for their 
contribution to the excellent performance of ALEPH. Those of us from non-member states wish to 
thank CERN for its hospitality.

\end{document}